\documentclass[usenatbib,usegraphicx]{mn2e}
\def\msun{{\rm M}_{\odot}}
\def\rsun{{\rm R}_{\odot}}

\begin{document}

\title[Rotation of low-mass stars in NGC 2516]{The Monitor project:
    Rotation of low-mass stars in the open cluster NGC 2516}
\author[J.~M.~Irwin et al.]{Jonathan~Irwin$^{1}$, Simon~Hodgkin$^{1}$,
    Suzanne~Aigrain$^{1}$, Leslie~Hebb$^{2}$
\newauthor
Jerome~Bouvier$^{3}$, Cathie~Clarke$^{1}$, Estelle~Moraux$^{3}$,
D.M.~Bramich$^{4}$ \\
$^{1}$Institute of Astronomy, University of Cambridge, Madingley Road,
  Cambridge, CB3 0HA, United Kingdom \\
$^{2}$School of Physics and Astronomy, University of St Andrews,
  North Haugh, St Andrews, KY16 9SS, Scotland \\
$^{3}$Laboratoire d'Astrophysique, Observatoire de Grenoble, BP 53,
  F-38041 Grenoble C\'{e}dex 9, France \\
$^{4}$Isaac Newton Group of Telescopes, Apartado de Correos 321,
  E-38700 Santa Cruz de la Palma, Canary Islands, Spain}
\date{Accepted .... Received ...; in original form ...}

\maketitle

\begin{abstract}
We report on the results of an $i$-band time-series photometric
survey of NGC 2516 using the CTIO 4m Blanco telescope and 8k Mosaic-II
detector, achieving better than $1\%$ photometric precision per data
point over $15 \la i \la 19$.  Candidate cluster members were selected
from a $V$ vs $V-I$ colour magnitude diagram over $16 < V < 26$
(covering masses from $0.7\ \msun$ down to below the brown dwarf
limit), finding $1685$ candidates, of which we expect $\sim 1000$ to
be real cluster members, taking into account contamination from the
field (which is most severe at the extremes of our mass range).
Searching for periodic variations in these gave $362$ detections over
the mass range $0.15 \la M/\msun \la 0.7$.  The rotation period
distributions were found to show a remarkable morphology as a function
of mass, with the fastest rotators bounded by $P > 0.25\ {\rm days}$,
and the slowest rotators for $M \la 0.5\ \msun$ bounded by a line of $P
\propto M^3$, with those for $M \ga 0.5\ \msun$ following a flatter
relation closer to $P \sim {\rm constant}$.  Models of the rotational
evolution were investigated, finding that the evolution of the fastest
rotators was well-reproduced by a conventional solid body model with
a mass-dependent saturation velocity, whereas core-envelope decoupling
was needed to reproduce the evolution of the slowest rotators.  None
of our models were able to simultaneously reproduce the behaviour of
both populations.
\end{abstract}

\section{Introduction}
\label{intro_section}

NGC 2516 is a well-studied nearby intermediate-age open cluster ($\sim
150\ {\rm Myr}$; \citealt*{jth2001}), at a distance modulus $(M - m)_0
= 7.93 \pm 0.14$ \citep{ter2002}.  The canonical reddening along the
line of sight to the cluster is $E(B - V) = 0.12$, implying a
photometrically derived metallicity of ${\rm [Fe/H]} = -0.05 \pm 0.14$
\citep{ter2002}.  The same authors derive a spectroscopic metallicity
from two stars of ${\rm [Fe/H]} = 0.01 \pm 0.03$, consistent within
the errors, and implying a slightly larger distance modulus $(M - m)_0
= 8.05 \pm 0.11$.  We have adopted the smaller distance modulus of $(M
- m)_0 = 7.93 \pm 0.14$ and photometric metallicity for the remainder
of this work, but any differences induced in the results by the
uncertainty in distance modulus are nevertheless small.

\begin{figure}
\centering
\includegraphics[angle=270,width=3.2in]{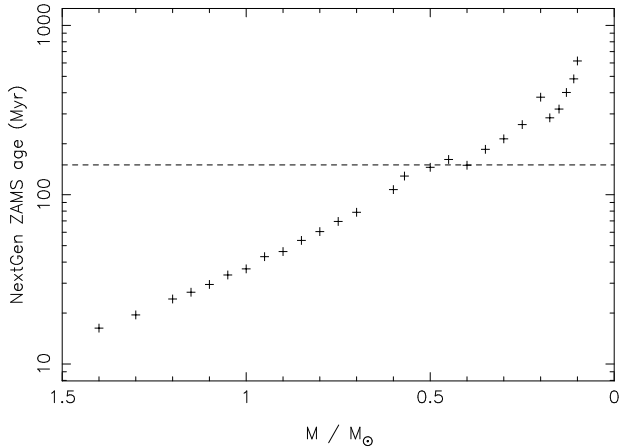}

\caption{Plot of the approximate age at which the ZAMS is reached for
  solar metallicity stars, from the NextGen models \citep{bcah98},
  computed as the time taken for the star to reach within $1 \%$ of
  its $1\ {\rm Gyr}$ age radius.  The horizontal dashed line indicates
  the approximate $150\ {\rm Myr}$ age of NGC 2516.}

\label{zamstime}
\end{figure}

NGC 2516 has often been referred to as the `Southern Pleiades' owing
to its similarity both in angular extent and age, where we adopt
an age of $100\ {\rm Myr}$ \citep{mmm93} for the Pleiades, using the
classical main sequence turnoff ages throughout this work for
consistency with the estimate for NGC 2516.  The Pleiades metallicity
is very similar, at ${\rm [Fe/H]} = -0.034 \pm 0.024$ \citep{bf1990}.
Clusters in the age range between $\alpha$ Persei ($\sim 50\ {\rm
  Myr}$, e.g. \citealt{bm99}) and the Hyades ($\sim 625\ {\rm Myr}$,
\citealt{p98}) are important for constraining stellar evolution
models.  Figure \ref{zamstime} shows a plot of the age at which a
solar metallicity star reaches the zero age main sequence (ZAMS) as a
function of stellar mass, indicating that solar mass stars reach the
ZAMS at $\sim 40\ {\rm   Myr}$, and $\sim 0.4-0.5\ \msun$ stars reach
the ZAMS at the age of NGC 2516.  It is therefore highly useful in
this regard owing to both its age and relative proximity, where we can
study solar mass stars after evolving for a short time on the main
sequence, and lower mass stars as they finish pre main sequence
evolution.  The similarity in age to the Pleiades (modulo the
uncertainty in this parameter) means that comparisons of these
clusters could be used to study the dependence of stellar parameters
on environment.

NGC 2516 is very well-studied in the literature.  The most relevant
existing work includes membership surveys (\citealt{jth2001},
\citealt{sung2002}), characterisation of the mass function
from solar mass \citep{jth2001} down to the brown dwarf regime
\citep{moraux2005}, a survey for radial velocities and binarity at the
bright end of the cluster sequence \citep{gonzalez2000}, and studies
of rotation using $v \sin i$ covering stars down to G spectral types
(\citealt{ter2002} and \citealt*{jjt1998}).  A number of X-ray
surveys, including monitoring for variability, have also been carried
out (e.g. \citealt{w2004}, \citealt{rhk2003}, \citealt{d2003},
\citealt{s2001}), the most recent of which \citep{pill2006} used a
deep {\it XMM-Newton} observation to probe the low-mass members (down
to $\sim$ M5) of the cluster.  They found that NGC 2516 stars are less
luminous in X-rays than the Pleiades members, and suggest that this
may be attributed to slightly older age, or lower rotation rates.

\subsection{Evolution of stellar angular momentum}
\label{amevol_section}

For a discussion of the context of this work, the reader is referred
to our M34 publication in \citet{i2006}.  The present survey probes a
different open cluster of slightly younger age ($\sim 150\ {\rm Myr}$
for NGC 2516, compared to $\sim 200\ {\rm Myr}$ for M34).  Due to the
small distance modulus and our use of a larger telescope, the NGC 2516
survey probes lower masses, covering a range of $0.1 \la M/\msun \la
0.7$, compared to $0.25 \la M/\msun \la 1.0$ in M34, allowing the very
low-mass stars to be examined in much greater detail.  This will allow
models of angular momentum evolution to be tested over a wider range
of stellar mass, in particular to determine whether models
incorporating solid body rotation (e.g. \citealt*{bfa97}) are able to
reproduce the  observed low-mass angular momentum evolution.  At the
lowest masses, core-envelope decoupling should not be relevant, since
all the stars will be fully-convective and thus rotate approximately
as solid bodies, allowing us to constrain directly the angular
momentum loss law without needing to consider the details of internal
angular momentum transport.  At the age of NGC 2516, stars with masses
$M \la 0.4\ \msun$ should be fully-convective \citep{cb97}.

\subsection{The survey}

We have undertaken a photometric survey in NGC 2516 using the CTIO 4m
Blanco telescope and Mosaic-II imager.  Our goals are two-fold: first,
to study rotation periods for a sample of low-mass members, covering M
spectral types, down to $\sim 0.1\ \msun$, and second, to look for
eclipsing binary systems containing low-mass stars, to obtain
dynamical mass measurements in conjunction with radial velocities from
follow-up spectroscopy.  Such systems provide the most accurate
determinations of fundamental stellar parameters (in particular,
masses) for input to models of stellar evolution, which are poorly
constrained in this age range.  We defer discussion of our eclipsing
binary candidates to a later paper once we have obtained suitable
follow-up spectroscopy. 

These observations are part of a larger photometric monitoring
survey of young open clusters over a range of ages and metalicities
(the Monitor project; \citealt{hodg06} and \citealt{a2007}).

The remainder of the paper is structured as follows: the observations
and data reduction are described in \S \ref{odr_section}, and the
colour magnitude diagram (CMD) of the cluster and candidate membership
selection are presented in \S \ref{memb_section}.  The method we use
for obtaining photometric periods is summarised in \S
\ref{period_section} (see \citealt{i2006} for a more detailed
discussion), and our results are summarised in \S
\ref{results_section}.  We discuss the implications of these results in
\S \ref{disc_section}, and our conclusions are summarised in \S
\ref{conclusions_section}. 

In several of the following sections, we use mass and radius estimates
for the cluster members.  We note here that these were derived from
the $150\ {\rm Myr}$ NextGen models, using the $I$-band magnitudes,
rather than $V - I$ colour or $V$ magnitude, for reasons discussed in
\S \ref{cmd_section}.

\section{Observations and data reduction}
\label{odr_section}

Photometric monitoring observations were obtained using the 4m CTIO
Blanco telescope, with the Mosaic-II imager, during two four-night
observing runs, separated by $\sim 1\ {\rm week}$, in 2006
February-March.  This instrument provides a field of view of $\sim 36'
\times 36'$ ($0.37\ {\rm sq. deg}$), using a mosaic of eight ${\rm 2k}
\times {\rm 4k}$ pixel CCDs, at a scale of $\sim 0.27'' / {\rm pix}$.
NGC 2516 was observed for $\sim 8\ {\rm hours}$ per night.

\begin{figure}
\centering
\includegraphics[angle=0,width=3.2in]{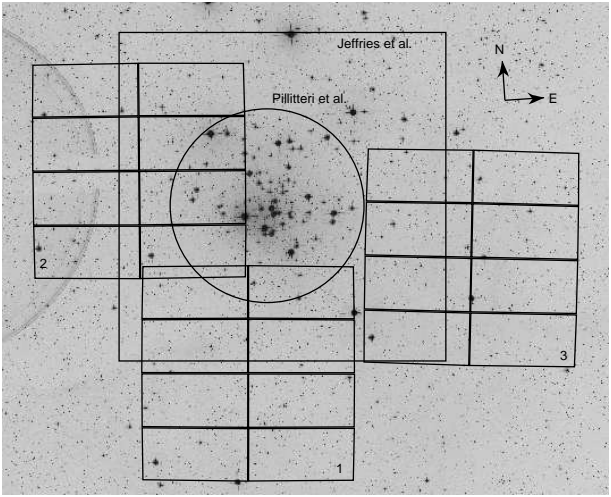}

\caption{DSS image of NGC 2516 covering $\sim 1.8^\circ \times
  1.5^\circ$, centred on {\tt 07 57 25 -60 55 00} (J2000), showing the
  coverage of the present survey (numbered 8-chip mosaic tiles), the
  XMM survey of \citet{pill2006}, and the optical survey of
  \citet{jth2001}.}

\label{coverage}
\end{figure}

In order to maximise the number of cluster members covered by our
survey, and given the large angular extent of NGC 2516, we used three
fields, with a single $75\ {\rm s}$ $i$-band exposure in each,
observed by cycling around all three fields, to give a cadence of
$\sim 9\ {\rm minutes}$, covering $\sim 1\ {\rm sq. deg}$ of the
cluster (illustrated in Figure \ref{coverage}).  Our fields were
chosen to be outside the central few arcmin of the cluster to avoid
contamination of the images from a number of very bright ($V \sim
5-8$) stars close to the cluster centre.

\begin{figure}
\centering
\includegraphics[angle=270,width=3.2in]{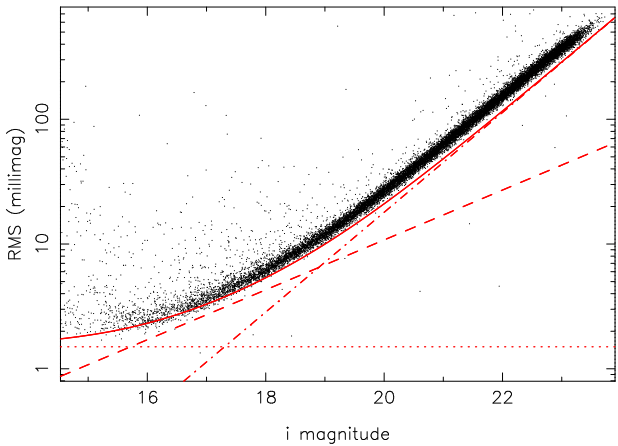}

\caption{Plot of RMS scatter as a function of magnitude for the
$i$-band observations of a single field in NGC 2516, for all unblended
objects with stellar morphological classifications.  The diagonal
dashed line shows the expected RMS from Poisson noise in the object,
the diagonal dot-dashed line shows the RMS from sky noise in the
photometric aperture, and the dotted line shows an additional $1.5\
{\rm mmag}$ contribution added in quadrature to account for systematic
effects.  The solid line shows the overall predicted RMS, combining
these contributions.}

\label{rmsplot}
\end{figure}

The observing conditions were superb, with exceptional seeing ($<
0.8''$ on a few nights) and $\sim 6$ photometric nights.  We also
obtained a number of longer $V$-band exposures in photometric
conditions ($3 \times$ $300{\rm s}$ in each field) to generate a
colour magnitude diagram (CMD).  Our observations are sufficient to
give $1 \%$ or better photometric precision pre data point from
saturation at $i \sim 15$ down to $i \sim 19$ (see Figure
\ref{rmsplot}), covering early to mid M spectral types at the age and
distance of NGC 2516.

For a full description of our data reduction steps, the reader is
referred to \citet{i2007}.  Briefly, we used the pipeline for
the INT wide-field survey \citep{il2001} for 2-D 
instrumental signature removal (crosstalk correction, bias correction,
flatfielding) and astrometric and photometric calibration.  We then
generated the {\em master catalogue} for each filter by stacking $20$
of the frames taken in the best conditions (seeing, sky brightness and
transparency) and running the source detection software on the stacked
image.  The resulting source positions were used to perform aperture
photometry on all of the time-series images.  We achieved a per data
point photometric precision of $\sim 2-4\ {\rm mmag}$ for the
brightest objects, with RMS scatter $< 1 \%$ for $i \la 19$ (see
Figure \ref{rmsplot}).

Our source detection software flags as likely blends any objects
detected as having overlapping isophotes.  This information is used,
in conjunction with a morphological image classification flag also
generated by the pipeline software \citep{il2001} to allow us to
identify non-stellar or blended objects in the time-series
photometry.

Photometric calibration of our data was carried out using regular
observations of \citet{l92} equatorial standard star fields in the
usual way.

Lightcurves were extracted from the data for $\sim 100\,000$ objects,
$63\,000$ of which had stellar morphological classifications, using our
standard aperture photometry techniques, described in \citet{i2007}.
We fit a 2-D quadratic polynomial to the residuals in each frame 
(measured for each object as the difference between its magnitude on
the frame in question and the median calculated across all frames) as
a function of position, for each of the $8$ CCDs separately.
Subsequent removal of this function accounted for effects such as
varying differential atmospheric extinction across each frame.  Over a
single CCD, the spatially-varying part of the correction remains
small, typically $\sim 0.02\ {\rm mag}$ peak-to-peak.  The reasons for
using this technique are discussed in more detail in \citet{i2007}.

For the production of deep CMDs, we stacked 20 $i$-band observations,
taken in good seeing and photometric conditions.  The limiting
magnitudes, measured as the approximate
magnitude at which our catalogues are $50\%$ complete (see \S
\ref{comp_section}) on these images were $V \simeq 24.3$ and $i
\simeq 22.7$.

\section{Selection of candidate low-mass members}
\label{memb_section}

Catalogues of photometrically-selected candidate members were
available from \citet{jth2001} for solar mass down to $\sim 0.2\
\msun$ and for $\sim 0.2 - 0.05\ \msun$ from \citet{moraux2005}, but
we elected to perform a new photometric selection using $V$ versus
$V-I$ CMDs from our data, to match exactly the field of view and
magnitude range of our time-series photometry.  Unfortunately, since
we have avoided the very centre of the cluster in our field selection
to reduce the effect of saturated bright stars on our CCD frames,
the overlap in sky coverage between the present survey and many of the
previous ones (e.g. \citealt{jth2001} and \citealt{pill2006}) is
relatively poor.

\subsection{The $V$ versus $V - I$ CMD}
\label{cmd_section}

\begin{figure}
\centering
\includegraphics[angle=270,width=3.5in]{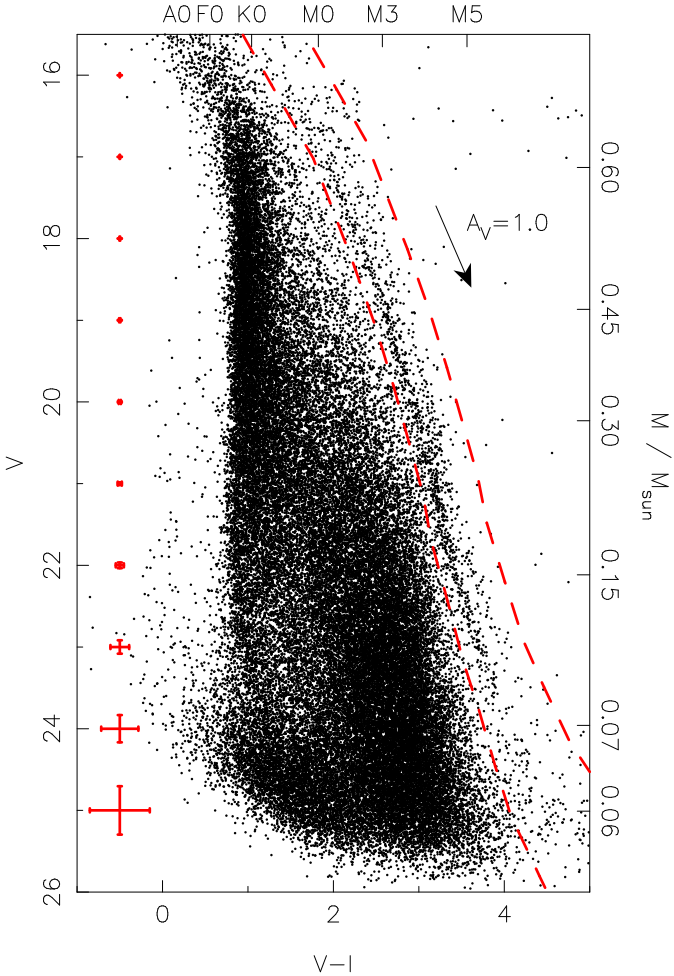}

\caption{$V$ versus $V - I$ CMD of NGC 2516 from stacked images, for
all objects with stellar morphological classification.  The cluster
main sequence is clearly visible on the right-hand side of the diagram.
The boundaries of the region to select photometric candidate members
are shown by the dashed lines (all objects between the dashed lines
were selected).  The reddening vector for $A_V = 1.0$ is shown at the
right-hand side of the diagram.  The mass scale is from the $150\ {\rm
  Myr}$ NextGen models \citep{bcah98} for $M > 0.1\ \msun$, and the
$120\ {\rm Myr}$ DUSTY models \citep{cbah2000} for $M < 0.1\ \msun$,
using our empirical isochrone to convert the $V$ magnitudes to $I$
magnitudes, and subsequently obtaining the masses from these, due to
known problems with the $V$ magnitudes from the models (see \S
\ref{cmd_section}).  The error bars at the left-hand side of the plot
indicate the typical photometric error for an object on the cluster
sequence.}

\label{cmd}
\end{figure}

Our CMD of NGC 2516 is shown in Figure \ref{cmd}.  The $V$ and $i$
measurements were converted to the standard Johnson-Cousins
photometric system using colour equations derived from our standard
star observations:
\begin{eqnarray}
(V - I)& = &(V_{ccd} - i_{ccd})\ /\ 0.899 \\
V& = &V_{ccd} + 0.005\ (V - I) \\
I& = &i_{ccd} - 0.096\ (V - I)
\end{eqnarray}

Candidate cluster members were selected by defining an empirical main
sequence `by eye' to follow the clearly-visible cluster single-star
sequence.  The cuts were defined by moving this line along a vector
perpendicular to the cluster sequence, by amounts $k - \sigma(V -
I)$ and $k + \sigma(V - I)$ as measured along this vector, where
$\sigma(V - I)$ is the photometric error in the $V - I$ colour.  The
values of $k$ used were $-0.15\ {\rm mag}$ for the lower line and
$0.5\ {\rm mag}$ for the upper line on the diagram, making the
brighter region wider to avoid rejecting binary and multiple systems,
which are overluminous for their colour compared to single stars.
$1685$ candidate photometric members were selected, over the full $V$
magnitude range from $V = 15.5$ to $26$, but the well-defined main
sequence appears to terminate at the hydrogen burning mass limit of $M
\sim 0.072\ \msun$, or $V \sim 24$, with a few candidate brown dwarfs
found below this limit, but with high field contamination.

We also considered using the model isochrones of \citet{bcah98} and
\citet{cbah2000} for selecting candidate members.  The NextGen model
isochrones were found to be unsuitable due to the known discrepancy
between these models and observations in the $V - I$ colour for
$T_{\rm eff} \la 3700\ {\rm K}$ (corresponding here to $V - I \ga 2$).
This was examined in more detail by \citet{bcah98}, and is due to a
missing source of opacity at these temperatures, leading to
overestimation of the $V$-band flux.  Consequently, when we have used
the NextGen isochrones to determine model masses and radii for our
objects, the $I$-band absolute magnitudes were used to perform the
relevant look-up, since these are less susceptible to the missing
source of opacity, and hence give more robust estimates.

\subsection{Completeness}
\label{comp_section}

\begin{figure}
\centering
\includegraphics[angle=270,width=3in]{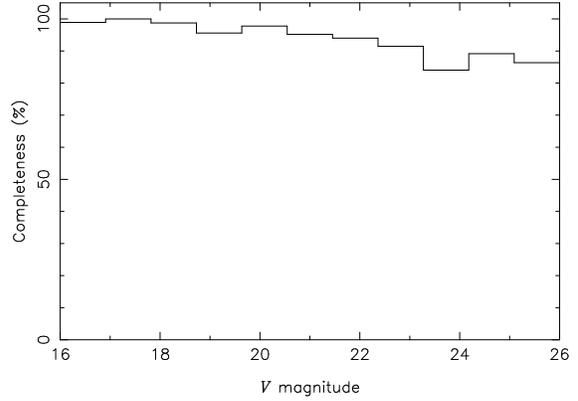}

\caption{Completeness in our source detection, measured as the
  fraction of simulated objects which were detected in each magnitude
  bin.  The diagram is plotted in the Johnson $V$ system assuming
  objects to lie on the empirically-derived cluster main sequence
  track of \S \ref{cmd_section} and Figure \ref{cmd}.  The diagram has
  been terminated at $V = 26$ since we have not continued the
  selection of cluster members past this magnitude due to termination
  of the apparent main sequence in the CMD of Figure \ref{cmd}.}

\label{comp}
\end{figure}

The completeness of our source detection was estimated by inserting
simulated stars at random $x,y$ positions into our images, drawing the
stellar magnitudes from a uniform distribution.  Figure \ref{comp}
shows the resulting plot of completeness as a function of $V$-band
magnitude.  The completeness for objects on the cluster sequence
is $> 90 \%$ down to the BD limit ($V \sim 24$).  We suspect
that the gradual drop in completeness toward fainter magnitudes in
Figure \ref{comp} results from a variable background in our CCD images
caused by halos and scattered light from the many bright stars in the
cluster field, impeding detection of faint sources in certain
regions of the frame.

\subsection{Contamination}
\label{contam_section}

\begin{figure}
\centering
\includegraphics[angle=270,width=3in]{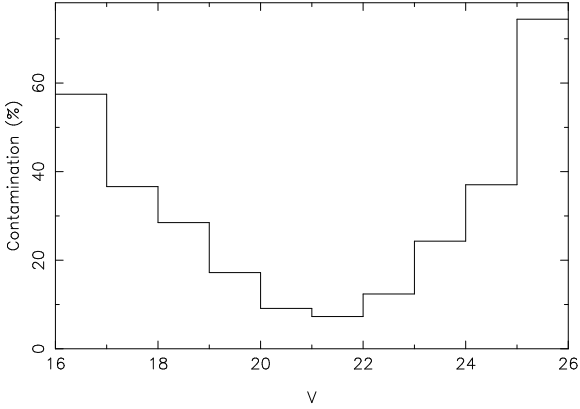}

\caption{Contamination, measured as the ratio of the calculated number
  of objects in each magnitude bin from the Galactic models, to the
  number of objects detected and classified as candidate cluster
  members in that magnitude bin.}

\label{contam}
\end{figure}

In order to estimate the level of contamination in our catalogue, we
used the Besan\c{c}on Galactic models \citep{r2003} to generate a
simulated catalogue of objects passing our selection criteria at the
Galactic coordinates of NGC 2516 ($l = 273.8^\circ$, $b = -15.9^\circ$),
covering the total FoV of $\sim 1.1\ {\rm sq.deg}$ (including gaps
between detectors).  We selected all objects over the apparent
magnitude range $14 < V < 26$, giving $105\,000$ stars.  The same
selection process as above for the cluster members was then applied to
find the contaminant objects.  A total of $651$ simulated objects
passed these membership selection criteria, giving an overall
contamination level of $\sim 39 \%$.  Figure \ref{contam} shows
the contamination as a function of $V$ magnitude.  We note that this
figure is somewhat uncertain due to the need to use Galactic models.
Follow-up data (spectroscopy) will be required to make a more
accurate estimate.  We note in particular that the apparently very
high contamination in the $V = 16-17$ bin, and possibly also the $V =
25-26$ bin, is probably the result of the model poorly-fitting the
observations, causing the contamination level to be overestimated.

\subsection{Binary fraction}
\label{binfrac_section}

\begin{figure}
\centering
\includegraphics[angle=270,width=3in]{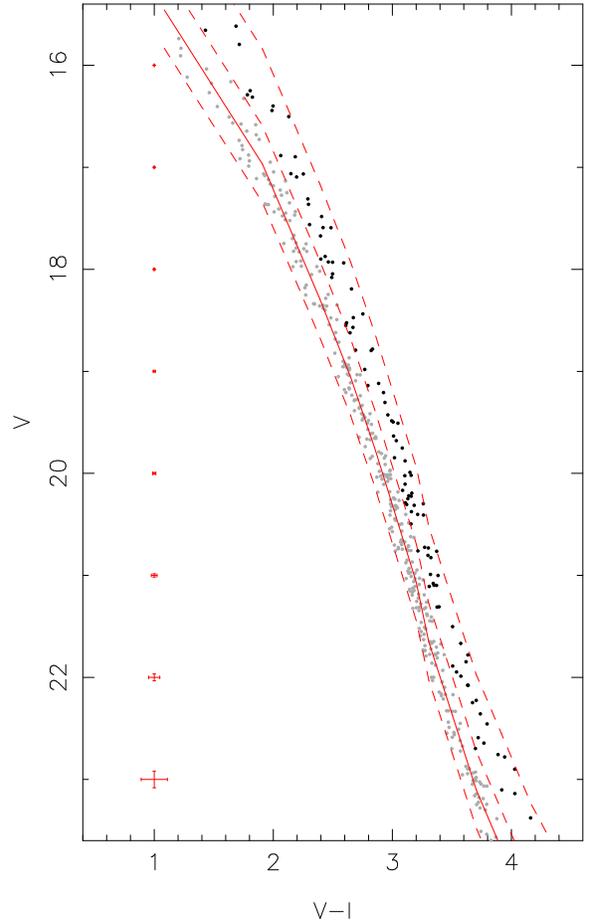}

\caption{$V$ versus $V - I$ CMD, as Figure \ref{cmd}, showing the
  selection of photometric candidate single stars (grey) and binaries
  (black).  The solid line shows the original empirical cluster main
  sequence, and the dashed lines show the boundaries used for
  selection, with $V$-band magnitudes $V + 1.25\ \log_{10} 2$, $V -
  1.25\ \log_{10} 2$, and $V - 3.75\ \log_{10} 2$, where $V$ is the
  empirical main sequence magnitude.  Typical photometric errors are
  indicated as for Figure \ref{cmd} by the error bars at the left-hand
  side of the diagram.}

\label{binary_diagram}
\end{figure}

Since the cluster binary sequence is so clearly visible in the CMD of
Figure \ref{cmd}, we have computed a simple photometric lower limit
for the binary fraction in the cluster as a function of mass, using
the empirical isochrone to select stars with luminosities lying within
a band of height $2.5 \log_{10} 2 \simeq 0.75\ {\rm mag}$ centred on the
single star cluster sequence, and the binary sequence (obtained by
shifting the cluster sequence by this amount to greater
luminosities), illustrated in Figure \ref{binary_diagram}.  This
technique suffers from several sources of systematic error, including:
(a) since the single star sequence is closer to the maximum of the
distribution of field stars in the CMD, the single star counts are
increased to a greater degree than the binary counts by field
contamination, (b) binaries with mass ratios significantly different
from $1$ lie closer to the single star sequence, and therefore tend
not to be selected as binaries, and (c) at the faint end of the CMD
the binaries are brighter, and hence more readily detected than single
stars of the same colour.

\begin{figure}
\centering
\includegraphics[angle=270,width=3in]{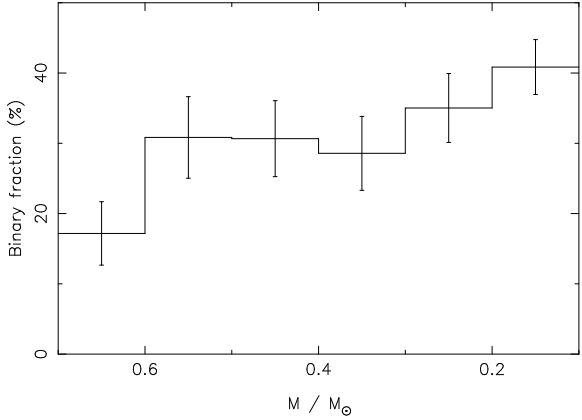}

\caption{Photometric binary fraction as a function of stellar mass
  (computed using the NextGen models of \citealt{bcah98}).}

\label{binaryfrac}
\end{figure}

The binary fraction is shown as a function of mass in Figure
\ref{binaryfrac}.  A total of $357$ candidate binaries and $1065$ 
candidate single stars were selected, giving an overall binary
fraction of $33.5 \pm 2.1\ \%$ (using Poisson errors).  

We have been unable to use the more traditional $I-K$ CMD approach
(e.g. \citealt{pin2003}) to determining limits on the binary fraction
due to the lack of suitable near-IR data.  2MASS provides $K$-band
photometry, but the resulting $I-K$ CMD did not show a clearly-visible
binary sequence due to the large scatter caused by the photometric
errors at the faint end of 2MASS.

At this point, it is prudent to examine the sensitivity of our
selection to the mass ratio $q$ of the binaries.  To do this, we have
simulated binaries over a range of $q$ from $0$ to $1.0$, using the
PMS models to compute the total $I$-band luminosity of the binary, and
our empirical isochrone to convert the individual $I$-band magnitudes
of the primary and secondary to $V$-band, and hence place the objects
on the $V-I$ CMD.  Our selection was then applied to determine the
minimum mass ratio $q$ for the detected binaries, defined as $q = M_2
/ M_1$ (where $M_1$ and $M_2$ are the primary and secondary masses,
respectively).  In general we find that our selection is sensitive to
smaller mass ratios at lower masses, which could explain the form of
the binary fraction as a function of mass seen in Figure
\ref{binaryfrac}.  The figure indicates that we are sensitive to $q
\ga 0.7$ over the entire mass range, and at best, $q \ga 0.3$ at $M
\sim 0.2\ \msun$.  Comparing these values to those of \citet{pin2003},
we find that the $V-I$ selection is slightly less sensitive to unequal
mass ratio systems (e.g. in their $0.6 - 1.0\ \msun$ bin the minimum
detectable mass ratio was $q = 0.5$ whereas here it is $q = 0.6$).

\begin{figure}
\centering
\includegraphics[angle=270,width=3in]{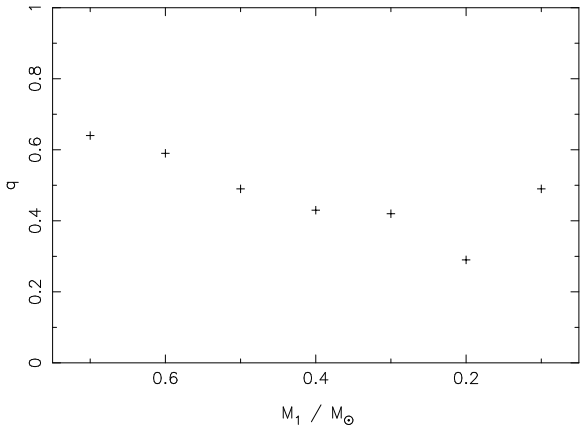}

\caption{Minimum mass ratio $q$ for detection of binaries as a
  function of the primary mass $M_1$ using our CMD approach.  Computed
  using the NextGen (for $M > 0.1\ \msun$) and DUSTY \citep{cbah2000}
  models (for $M \le 0.1\ \msun$) to convert between mass and $I$-band
  absolute magnitude, and our empirical main sequence isochrone to
  convert between $I$-band and $V$-band magnitudes.  The curve is not
  smooth owing to the coarse binning used in our empirical isochrone.}

\label{minq}
\end{figure}

We note that since the completeness in Figure \ref{comp} is close to
constant over the entire mass range covered by Figure
\ref{binaryfrac}, the apparent increase in binary fraction toward
lower masses is probably not an incompleteness effect resulting from
lower completeness for the single star sequence, compared to the more
luminous binary sequence.  However, our selection is affected by field
contamination, with a background of field objects falling inside the
selection bins on the CMD.  This will tend to decrease the computed
binary fraction, since there are more field objects at the single star
side of the selection region (see Figure \ref{cmd}).  It is difficult
to correct for this effect until we have obtained follow-up
spectroscopy.

Our results in $V-I$ are in good general agreement with those of
\citet{pin2003}, although it is difficult to distinguish whether our
data are consistent with any change in the binary fraction as a
function of mass, given the small number statistics (and hence large
Poisson errors in Figure \ref{binaryfrac}) and better sensitivity to
small $q$ with decreasing mass (leading presumably to more detections).

Photometric selection of binaries remains, to our knowledge,
relatively untested, without detailed evaluation of the sensitivity.
Therefore, a more detailed and reliable study of binarity in NGC 2516
requires follow-up spectroscopy.

\section{Period detection}
\label{period_section}

\subsection{Method}
\label{method_section}

The method we use for detection of periodic variables is described in
detail in \citet{i2006}, and we provide only a brief
summary here.  The method uses least-squares fitting of sine curves to
the time series $m(t)$ (in magnitudes) for {\em all} candidate cluster
members, using the form:
\begin{equation}
m(t) = m_{dc} + \alpha \sin(\omega t + \phi)
\label{sine_eqn}
\end{equation}
where $m_{dc}$, $\alpha$ (the amplitude) and $\phi$ (the phase) are
free parameters at each value of $\omega$ over an equally-spaced grid
of frequencies, corresponding to periods from $0.005 - 50\ {\rm days}$
for the present data-set.

Periodic variable lightcurves were selected by evaluating the change
in reduced $\chi^2$:
\begin{equation}
\Delta \chi^2_\nu = \chi^2_\nu - \chi^2_{\nu,smooth} > 0.4
\end{equation}
where $\chi^2_\nu$ is the reduced $\chi^2$ of the original lightcurve
with respect to a constant model, and $\chi^2_{\nu,smooth}$ is the
reduced $\chi^2$ of the lightcurve with the smoothed, phase-folded
version subtracted.  This threshold was used for the M34 data and
appears to work well here too, carefully checked by examining all the
lightcurves for two of the detectors, chosen randomly.  A total of
$1011$ objects were selected by this automated part of the procedure.

The selected lightcurves were examined by eye, to define the final
sample of periodic variables.  A total of $362$ lightcurves were
selected, with the remainder appearing non-variable or too ambiguous
to be included.

\subsection{Simulations}
\label{sim_section}

Monte Carlo simulations were performed following the method detailed
in \citet{i2006}, injecting simulated signals of $2 \%$ amplitude and
periods chosen following a uniform distribution on $\log_{10}$ period
from $0.1$ to $20\ {\rm days}$, into lightcurves covering a uniform
distribution in mass, from $0.7$ to $0.1\ \msun$.  A total of $2271$
objects were simulated.

\begin{figure*}
\centering

\includegraphics[angle=270,width=6in]{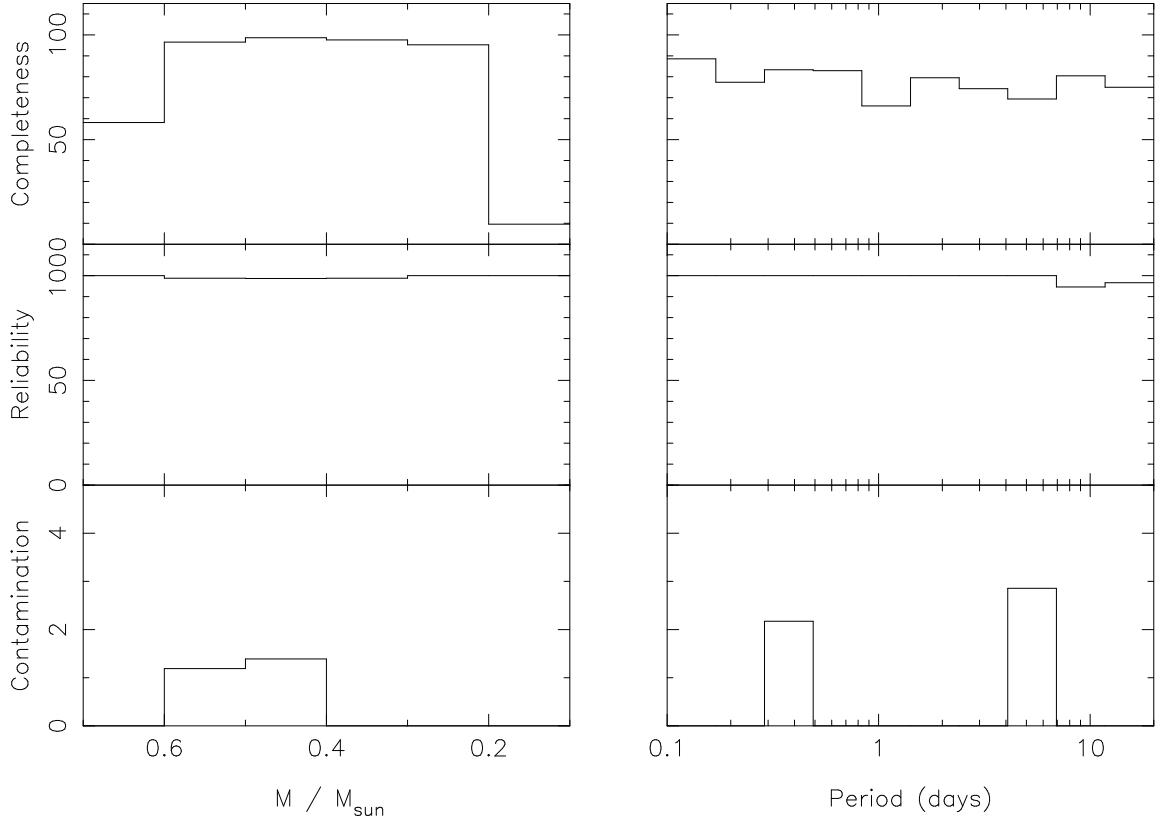}

\caption{Results of the simulations for $0.02\ {\rm mag}$ amplitude,
  plotted as a function of mass (left) and period (right).  The
  simulated region covered $0.1 < {\rm M}/\msun < 0.7$ in 
  order to be consistent with the NGC 2516 sample.  {\bf Top panels}:
  completeness as a function of real (input) period.  {\bf Centre
  panels}: Reliability of period determination, plotted as the fraction
  of objects with a given true period, detected with the correct
  period (defined as differing by $< 20\%$ from the true period).
  {\bf Bottom panels}: Contamination, plotted as the fraction of
  objects with a given detected period, having a true period differing
  by $> 20\%$ from the detected value.}

\label{sim_results}
\end{figure*}

\begin{figure}
\centering
\includegraphics[angle=270,width=3in,bb=59 107 581 630,clip]{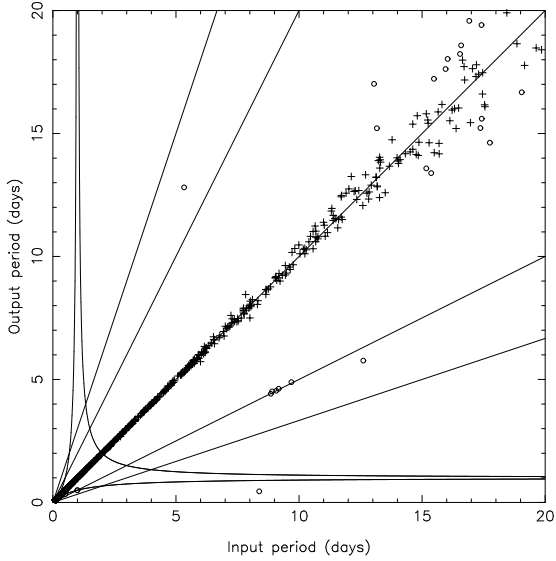}

\caption{Detected period as a function of actual (input) period for our
  simulations.  Objects plotted with crosses had fractional period
  error $< 10\%$, open circles $> 10\%$.  The straight lines represent
  equal input and output periods, and factors of $2$, $3$, $1/2$ and
  $1/3$.  The curved lines are the loci of the $\pm 1\ {\rm day^{-1}}$
  aliases resulting from gaps during the day.  The majority of the
  points fall on (or close to) the line of equal periods.}

\label{periodcomp}
\end{figure}

The results of the simulations are shown in Figure \ref{sim_results}
as greyscale diagrams of completeness, reliability and contamination
as a function of period and stellar mass.  Broadly, our period
detections are close to $100 \%$ complete down to $0.2\ \msun$, with
remarkably little period dependence.  The slight incompleteness
(approximately $63 \%$ completeness for $0.6 < M/\msun < 0.7$) seen
in the highest mass bin is a result of saturation.  Figure
\ref{periodcomp} shows a comparison of the detected periods with real
periods for our simulated objects, indicating remarkably high
reliability, especially compared to the earlier M34 data-set
\citep{i2006}, resulting from better sampling.

\subsection{Detection rate and reliability}

\begin{figure}
\centering
\includegraphics[angle=270,width=3.5in]{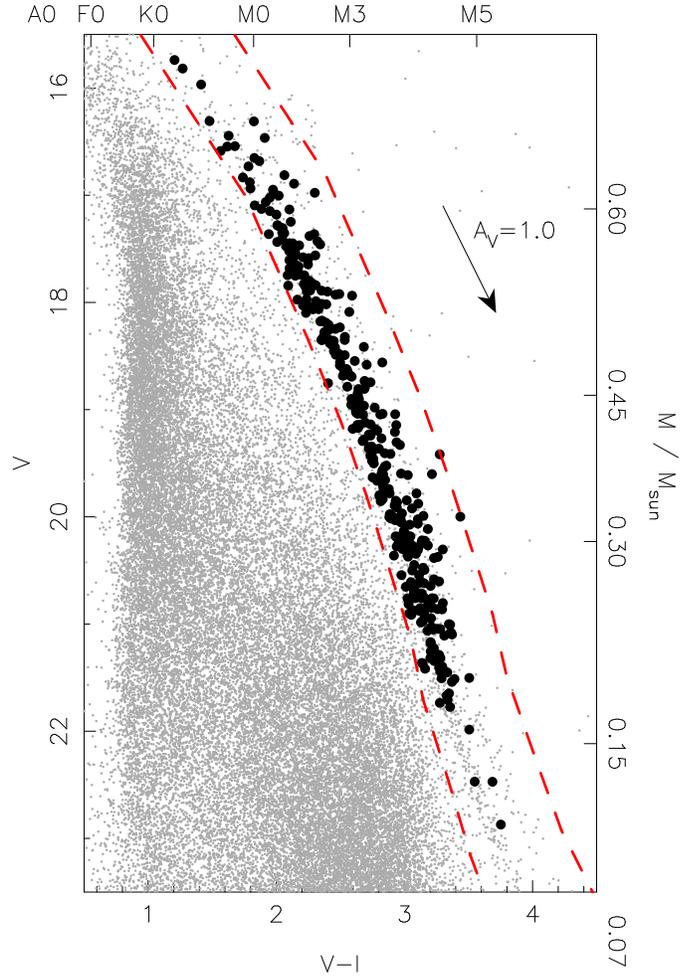}

\caption{Magnified $V$ versus $V - I$ CMD of NGC 2516, for objects
with stellar morphological classification, as Figure \ref{cmd},
showing all $362$ candidate cluster members with detected periods
(black points).  The dashed lines show the cuts used to select
candidate cluster members (see \S \ref{cmd_section}).}

\label{cands_on_cmd}
\end{figure}

The locations of our detected periodic variable candidate cluster
members on a $V$ versus $V-I$ CMD of NGC 2516 are shown in Figure
\ref{cands_on_cmd}.  The diagram indicates that the majority of
the detections lie on the single-star cluster main sequence, as would
be expected for rotation in cluster stars as opposed to, say,
eclipsing binaries.  The measured photometric binary fraction (see \S
\ref{binfrac_section}) for the periodic variable stars is
$26.3 \pm 3.5 \%$ (using Poisson errors).  This value is slightly
lower than the value for all cluster members from \S
\ref{binfrac_section}, but only at the $1.8 \sigma$ level.  Given
the systematic uncertainties in the binary fractions (which are not
accounted for by the Poisson counting errors), this is probably not a
significant result.

\begin{figure}
\centering
\includegraphics[angle=270,width=3in]{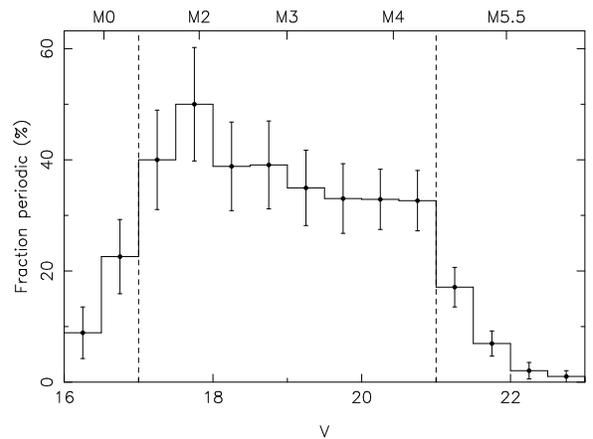}

\caption{Fraction of candidate cluster members detected as periodic
  variables, plotted as a function of magnitude.  This distribution
  has not been corrected for incompleteness in the period detections,
  which are close to $100\%$ complete for $17 < V < 21$ (shown by the
  vertical dashed lines).}

\label{fracper}
\end{figure}

Figure \ref{fracper} shows the fraction of cluster members with
detected periods as a function of $V$ magnitude.  The decaying parts
of the histogram at the bright and faint ends may be caused by the
increased field contamination here (see Figure \ref{contam}), since we
expect field objects on average show less rotational modulation than
cluster objects, and/or incompleteness effects resulting from
saturation for $V \la 17$, and for $V \ga 21$, the gradual increase in
the minimum amplitude of variations we can detect (corresponding to
the reduction in sensitivity moving to fainter stars, see Figure
\ref{rmsplot}).

In the M34 survey of \citet{i2006}, we commented on a possible
increase in the fraction of photometric variables from $K$ to $M$
spectral types, subject to a large uncertainty due to small number
statistics.  Since the mass range covered by the present sample is
different, direct comparison is difficult, and there is no clear
evidence for such a trend in NGC 2516.  We further note that the
decline observed in M34 at faint magnitudes is probably indeed due to
incompleteness as stated by \citet{i2006}, since a similar trend is
also seen in the NGC 2516 data-set, displaced accordingly for the
differences in sensitivity and distance modulus.

We note that lower field contamination (as evaluated in \S
\ref{contam_section}) is expected in the rotation sample than in the
full candidate membership sample.  Typical field population ages for
the young disc of $3\ {\rm Gyr}$ (\citealt{may74}, \citealt{msr91})
imply slower rotation rates by factors of a few than cluster
members, reduced activity, implying smaller asymmetric components of
the spot coverage and hence lower photometric amplitudes (which
probably render many of them undetectable).

The properties of all our rotation candidates are listed in Table
\ref{cand_table}.

\begin{table*}
\centering
\begin{tabular}{lllrrrrrrrr}
\hline
Identifier     &RA    &Dec   &$V$   &$I$   &$P$    &$\alpha_i$ &$M$ &$R$ \\
               &J2000 &J2000 &mag   &mag   &days   &mag        &$\msun$ &$\rsun$ \\
\hline
N2516-1-1-351  &07 56 28.22 &-61 27 46.6 &18.99 &16.35 & 2.318 &0.016 &0.44 &0.41 \\
N2516-1-1-784  &07 56 43.99 &-61 30 51.5 &21.25 &18.05 & 0.649 &0.013 &0.20 &0.24 \\
N2516-1-1-881  &07 56 47.57 &-61 34 37.9 &17.80 &15.54 & 7.677 &0.010 &0.55 &0.51 \\
N2516-1-1-958  &07 56 49.99 &-61 32 19.9 &18.48 &15.99 & 6.291 &0.015 &0.49 &0.45 \\
N2516-1-1-1470 &07 57 08.92 &-61 29 18.6 &17.67 &15.49 & 8.803 &0.016 &0.56 &0.52 \\
\hline
\end{tabular}

\caption{Properties of our $362$ rotation candidates.
  The period $P$ in days, $i$-band amplitude $\alpha_i$ (units of
  magnitudes, in the instrumental bandpass), interpolated mass and
  radius (from the models of \citealt{bcah98}, derived using the $I$
  magnitudes) are given (where available).  Our identifiers are formed
  using a simple scheme of the cluster name, field number, CCD number
  and a running count of stars in each CCD, concatenated with dashes.
  The full table is available in the electronic edition.}
\label{cand_table}
\end{table*}

\subsection{Non-periodic objects}

The population of objects rejected by the period detection procedure
described in \S \ref{period_section} was examined, finding that the
most variable population of these lightcurves (which might correspond
to non-periodic or semi-periodic variability) was contaminated by a
small number of lightcurves ($\sim 50$) exhibiting various uncorrected
systematic effects, mostly seeing-correlated variations due to image
blending.  It is therefore difficult to quantify the level of
non-periodic or semi-periodic variability in NGC 2516 from our
data.  Qualitatively however, there appear to be very few of these
variables, and examining the lightcurves indicated only $\sim 10$
obvious cases, some of which resembled eclipses (either planetary
transits or eclipsing binaries), and will be the subject of a later
Monitor project paper.

\section{Results}
\label{results_section}

\subsection{NGC 2516 rotation periods}
\label{prv_section}

\begin{figure}
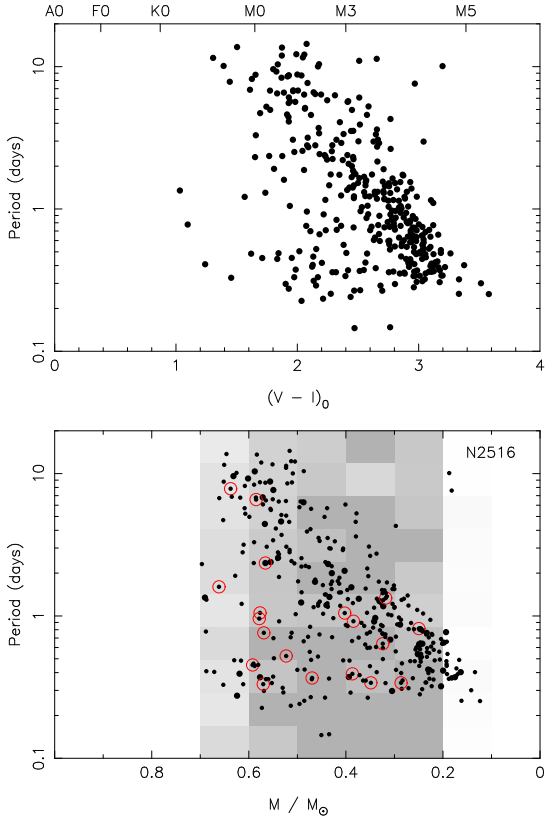

\centering
\includegraphics[angle=270,width=3in]{pcd_n2516.ps}
\includegraphics[angle=270,width=3in]{pmd_n2516.ps}

\caption{Plots of rotation period as a function of dereddened $V-I$
  colour (top), and mass (bottom) for NGC 2516, deriving the masses
  using the $150\ {\rm Myr}$ NextGen mass-magnitude relations of
  \citet{bcah98} and our measured $I$-band magnitudes.  In the lower
  diagram, the greyscales show the completeness for $0.02\ {\rm mag}$
  periodic variations from our simulations, and X-ray sources from
  \citet{pill2006} are overlaid with open circles.}

\label{pcd}
\end{figure}

Plots of period as a function of $V-I$ colour and mass for the objects
photometrically selected as possible cluster members are shown in
Figure \ref{pcd}.  These diagrams show a striking correlation between
stellar mass (or spectral type) and the longest rotation period seen
at that mass, with a clear lack of slow rotators at very low masses.
This trend is also followed by the majority of the rotators, with only
a tail of faster rotators to $\sim 0.25\ {\rm day}$ periods.
Furthermore, very few objects were found rotating faster than this,
implying a hard lower limit to the observed rotation periods at $0.25\
{\rm days}$.

\begin{figure}
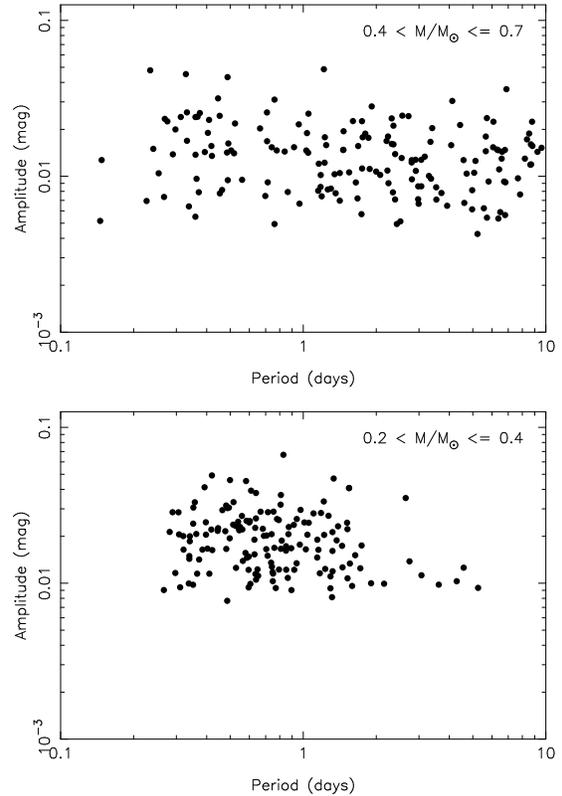

\centering
\includegraphics[angle=270,width=3in]{pad_n2516_high.ps}
\includegraphics[angle=270,width=3in]{pad_n2516_low.ps}

\caption{Plot of amplitude as a function of period for NGC 2516 in two
  mass bins: $0.4 < M/\msun < 0.7$ (top) and $0.2 < M/\msun < 0.4$
  (bottom).}

\label{pad}
\end{figure}

Could the apparent morphology in Figure \ref{pcd} be explained by
sample biases?  The simulations of \S \ref{sim_section} suggest that
this is unlikely, since we are sensitive to shorter periods than the
$0.2 - 0.3\ {\rm day}$ `limit', and the slight bias toward detection
of shorter periods at low mass is not sufficient to explain the
observations for the slow rotators.  Furthermore, Figure \ref{pad}
indicates that the lack of sensitivity to low amplitudes at low masses
does not appear to introduce any systematic changes in the detected
periods.

Figures \ref{pcd} and \ref{pad} show an apparent lack of objects with
masses $0.2 < M/\msun < 0.4$ and rotation periods $\ga 2\ {\rm days}$,
a region of the diagram where our survey is sensitive down to
amplitudes of $0.01\ {\rm mag}$ (see also \S \ref{sim_section}).
There is also an apparent lack of large-amplitude objects in this
region of the diagram, although this is not statistically significant
(applying a Kolmogorov-Smirnov test between the two mass bins
indicated a probability of $0.25$ that the distributions were drawn
from the same parent population).

\subsubsection{Period distributions}
\label{perioddist_section}

\begin{figure*}
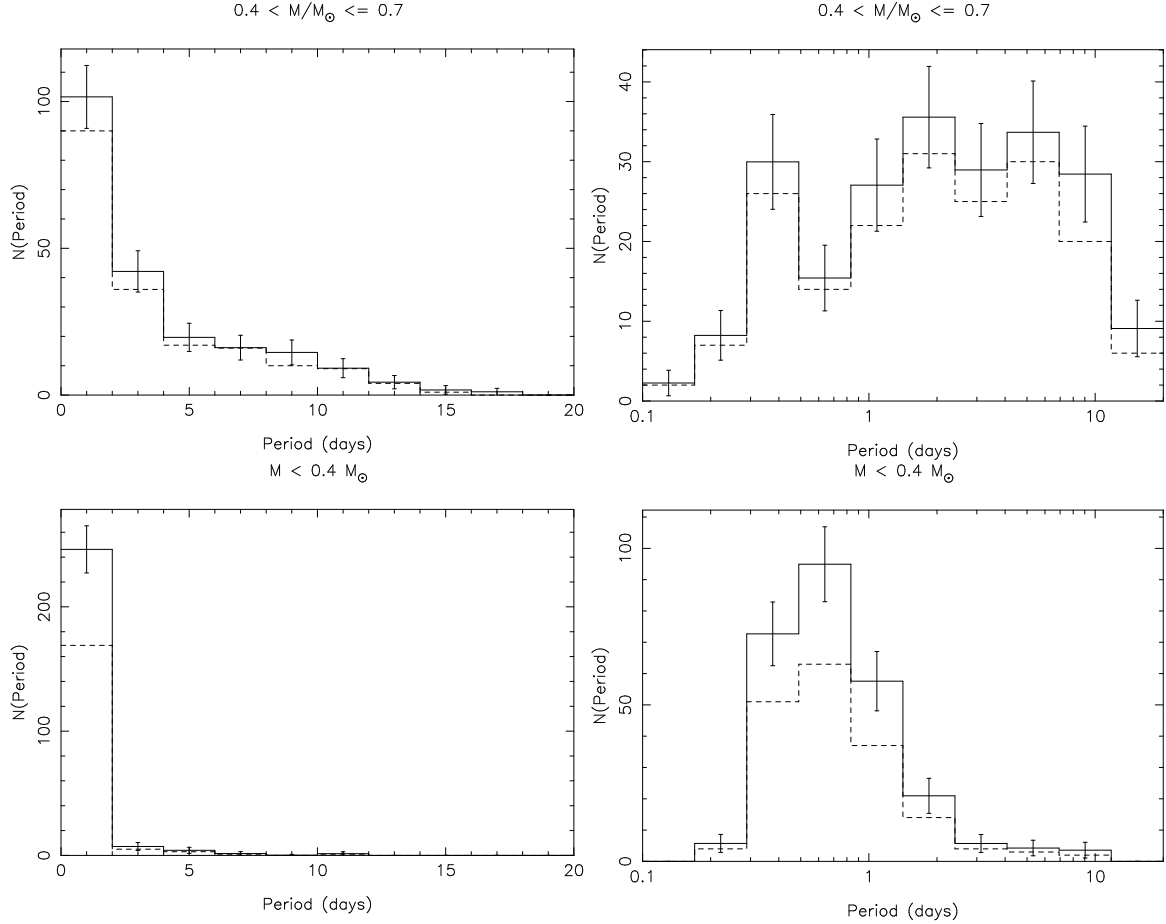

\centering
\includegraphics[angle=270,width=3in]{perioddist_1.ps}
\includegraphics[angle=270,width=3in]{perioddist_3.ps}
\includegraphics[angle=270,width=3in]{perioddist_2.ps}
\includegraphics[angle=270,width=3in]{perioddist_4.ps}

\caption{Period distributions for objects classified as possible
  photometric members, in two mass bins: $0.4 \le M/\msun < 0.7$
  (upper row, corresponding roughly to early-M spectral types) and $M
  < 0.4\ \msun$ (lower row, late-M).  The left-hand panels
  show the distributions plotted in linear period, and the right-hand
  panels show the same distributions plotted in $\log_{10}$ period.
  The dashed lines show the measured period distributions, and the
  solid lines show the results of attempting to correct for
  incompleteness and reliability, as described in the text.}

\label{perioddist}
\end{figure*}

In order to quantify the morphology of Figure \ref{pcd}, we have used
histograms of the rotation period distributions in two broad mass bins,
$0.4 \le M/\msun < 0.7$ and $M < 0.4\ \msun$, shown in Figure
\ref{perioddist}.  We have attempted to correct the distributions for
the effects of incompleteness and (un)reliability using the
simulations described in \S \ref{sim_section}, following the method
used in \citet{i2006}.  The results of doing this are shown in the
solid histograms in Figure \ref{perioddist}, and the raw period
distributions in the dashed histograms.

The period distributions in the two mass bins of Figure
\ref{perioddist} show clear differences, with the low-mass stars ($M <
0.4\ \msun$) showing a strongly peaked rotational period distribution,
with a maximum at $\sim 0.6-0.7\ {\rm days}$, whereas the higher-mass
stars ($0.4 \le M/\msun < 0.7$) show a broader distribution.  We
applied a two-sided Kolmogorov-Smirnov test to the corrected
distributions to confirm the statistical significance of this result,
finding a probability of $4 \times 10^{-20}$ that the distributions
were drawn from the same parent population.

The implication of this result is that the observed morphology in
Figure \ref{pcd}, and in particular the increase of the longest
observed rotation period as a function of increasing mass, a trend
followed also by the bulk of the rotators, is real and statistically
significant.

\subsubsection{Low-mass slow rotators}
\label{lmsr_section}

A group of $\sim 4$ outliers are visible e.g. in the upper panel of
Figure \ref{pcd}, as unusually slow rotators at low mass ($(V - I) \ga
2.5$).  These objects also give rise to the apparent `trend' seen in
Figure \ref{pad} of smaller amplitudes for periods $P \ga 2\ {\rm
  days}$, for the lower mass bin ($0.2 < M/\msun \le 0.4$).

Careful examination of the lightcurves indicates that these
objects are all convincing detections, and lie on, or close to, the
cluster sequence.  Obtaining optical spectroscopy for these objects
would appear to be warranted, to constrain their membership of the
cluster, and determine spectral types, to evaluate their reddening.

\subsubsection{Rapid rotators}

We have examined the periods of our fastest-rotating stars, to check
if they are rotating close to their break-up velocity.  The critical
period $P_{\rm crit}$ for break-up is given approximately by:
\begin{equation}
P_{\rm crit} = 0.116\ {\rm days}\ {(R / {\rm R_{\odot}})^{3/2}\over{(M / {\rm M_{\odot}})^{1/2}}}
\end{equation}
where $R$ and $M$ are the stellar radius and mass respectively
(e.g. \citealt{h2002}).  Using the NextGen models of \citet{bcah98},
the object rotating closest to breakup is N2516-2-8-624, with $P_{\rm
  crit} / P = 0.42$.  However, the majority of our objects are
rotating at much lower fractions of their break-up velocity.

Figure \ref{pad} indicates that all but three of our fastest rotators
have $P > 0.2\ {\rm days}$.  The outliers are N2516-1-4-3373 ($P =
0.148\ {\rm days}$), N2516-2-8-566 ($P = 0.146\ {\rm days}$) and
N2516-2-8-624 ($P = 0.087\ {\rm days}$).  Two of these objects,
N2516-1-4-3373 and N2516-2-8-624, clearly lie on the binary sequence
in Figure \ref{cmd}, and N2516-2-8-566 may also lie on the binary
sequence, so we suspect that these stars might be tidally locked
binaries.  They are therefore good candidates for spectroscopic
follow-up.

We note that there are three outliers visible in Figure \ref{pcd} with
$(V - I)_0 \la 1.3$, and periods close to $1\ {\rm day}$.  Similar
objects were seen in our M34 study \citep{i2006}.  At these masses,
field contamination in our survey is significant ($\sim 60\%$), so we
suspect that they may be contaminating field objects, and possibly
binaries (in M34, several of the corresponding objects were also
detected in X-rays, it is difficult to say if this holds in NGC 2516
due to the incomplete overlap in spatial coverage with the X-ray
catalogue of \citealt{pill2006}).  Our survey is incomplete at these
(high) masses due to saturation of the detectors, which varied between
CCDs in the mosaic, giving rise to the lack of other objects detected
in this range of $(V - I)_0$.

\subsection{Comparison with other data-sets}

\subsubsection{Period versus mass diagram}
\label{pmd_section}

\begin{figure}
\centering
\includegraphics[angle=270,width=3in]{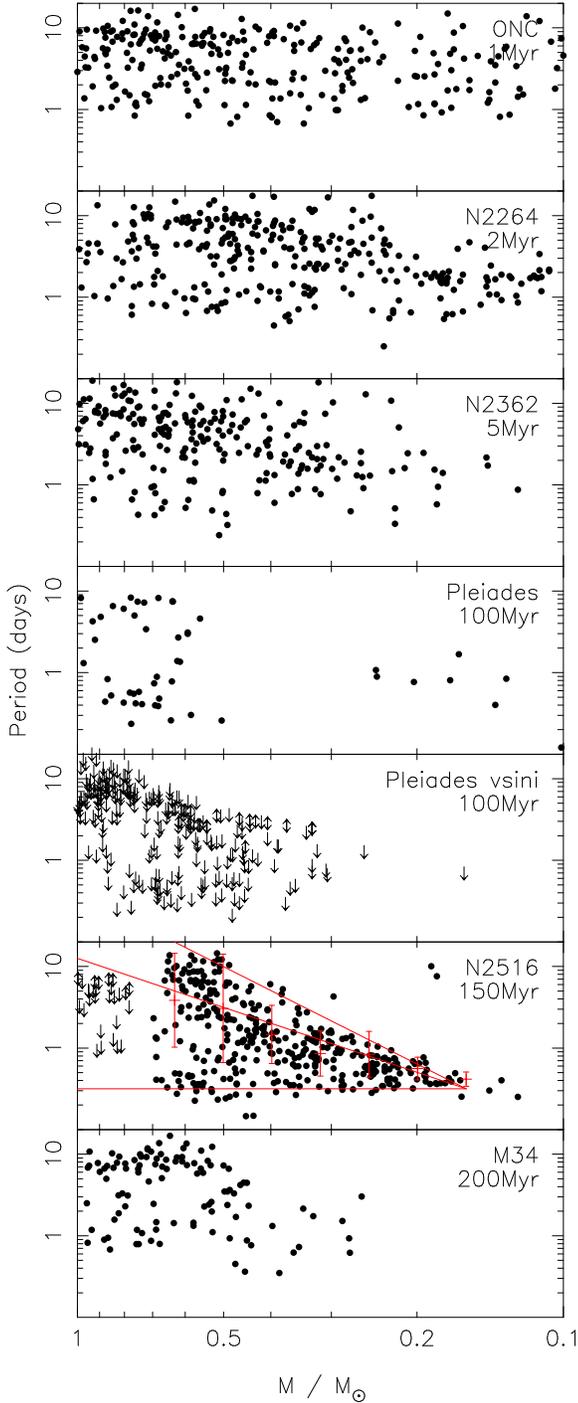}

\caption{Rotation period as a function of stellar mass for (top to
  bottom): ONC, NGC 2264, NGC 2362, the Pleiades (rotation
  periods and $v \sin i$ data on separate plots), NGC 2516 and M34.
  Lower and upper limits (from $v \sin i$ data) are marked with arrows.
  The masses were taken from the NextGen mass-magnitude relations
  \citep{bcah98} at the appropriate ages.  The ONC data are from
  \citet{h2002}.  For NGC 2264 we used the data of \citet{lamm05} and
  \citet{m2004}.  The NGC 2362 data are from the Monitor project, to
  be published in Hodgkin et al. (in prep).  The Pleiades rotation
  period data are a compilation of the results from \citet{ple1},
  \citet{ple2}, \citet{ple3}, \citet{ple4}, \citet{ple5},
  \citet{ple6}, \citet{ple7} (taken from the open cluster database),
  \citet{t99} and \citet{se2004b}.  The Pleiades $v \sin i$ data are a
  compilation of results from \citet{plev1}, \citet{plev2},
  \citet{plev3}, \citet{plev4}, \citet{plev5} and \citet{plev6}.  The
  M34 data are taken from \citet{i2006}.  In the NGC 2516 plot, the
  lines show $P = {\rm constant}$, $P \propto M^2$ and $P \propto
  M^3$.  The error bars show the median of the distribution binned in
  $0.1\ {\rm dex}$ bins of $\log M$.}

\label{pmd}
\end{figure}

Figure \ref{pmd} shows a diagram of rotation period as a
function of stellar mass for the ONC ($1 \pm 1\ {\rm Myr}$;
\citealt{h97}), NGC 2264 ($2-4\ {\rm Myr}$; \citealt{park2000}), NGC
2362 ($\sim 5 \pm 1\ {\rm Myr}$; \citealt{moi2001}, \citealt{bl96}),
the Pleiades ($\sim 100\ {\rm Myr}$; \citealt{mmm93}), NGC 2516 and
M34 ($\sim 200\ {\rm Myr}$; \citealt{jp96}).  Data sources for each
cluster are indicated in the figure caption.  From this Figure, it
appears that the objects we observe in NGC 2516 are bounded by lines
of $P \sim {\rm constant}$ (or a low power of $M$) for the fastest
rotators, and $P \propto M^3$ for the slow rotators in the mass range
$0.2 < M/\msun < 0.7$.

Furthermore, comparing the NGC 2516 and M34 results in Figure
\ref{pmd} indicates that the slowest rotators for $M > 0.6\ \msun$
have $P \sim {\rm constant}$ as a function of mass, whereas for $M <
0.6\ \msun$, the period increases as a function of mass.  A similar
trend is visible in the Pleiades $v \sin i$ data (noting that a number
of the slowest rotators here only have $v \sin i$ lower limits,
denoted by double-headed arrows).

The diagram clearly shows a gradual evolutionary sequence, from a
relatively flat mass-dependence of the rotation periods in the ONC
($\sim 1\ {\rm Myr}$), to a sloping relation in NGC 2362 ($\sim 5\
{\rm Myr}$), and the emergence of the break between a flat distribution
for $M \ga 0.6\ \msun$ and strongly sloping distribution at lower
masses, for the Pleiades, NGC 2516 ($\sim 150\ {\rm Myr}$) and M34
($\sim 200\ {\rm Myr}$).  We suggest this apparent change in slope at
$M \sim 0.6\ \msun$ may correspond to a transition to
convectively-dominated stellar interiors moving to lower masses (but
see \S \ref{model_section}).

\subsubsection{$v \sin i$ measurements}

\citet{ter2002} presented $v \sin i$ measurements for G and early-K
dwarfs in NGC 2516.  Dividing their sample into
branches of fast and slow rotation for $(B - V)_0 > 0.6$ and counting
objects with $v \sin i < 15\ {\rm km\ s^{-1}}$ as slow rotators, gave
$7$ fast rotators, with typical $v \sin i \sim 20-50\ {\rm km\
  s^{-1}}$, and $17$ slow rotators with $v \sin i \la 10\ {\rm km\
  s^{-1}}$, i.e. $30 \%$ are fast rotators.  The threshold of $v \sin i
= 15\ {\rm km\ s^{-1}}$ corresponds to $P = 2.2\ \sin i\ {\rm days}$
at $0.7\ \msun$, or $\langle P \rangle = 1.72\ {\rm days}$, using
$\langle \sin i \rangle = \pi / 4$ (averaged over all inclinations
$i$).  Applying this threshold to the $0.5 - 0.7\ \msun$ bin of Figure
\ref{pcd}, we counted $69$ slow rotators, and 40 fast rotators,
i.e. $37 \%$ fast rotators (and clearly from Figure \ref{pcd} this
fraction increases considerably for lower masses, to nearly $100 \%$
at $0.2\ \msun$).  Furthermore, for $(B - V)_0 < 0.6$ there are no
clearly-visible fast rotators in Figure 5 of \citet{ter2002}.  These
results clearly indicate that the fraction of objects rotating faster
than the chosen threshold increases as a function of mass, lending
support to the rotation period results we have presented.

\subsubsection{Comparison to the field: main sequence spin-down}

Comparing our results with the field survey of \citet{del98}
indicates that the spin-down timescale of M-dwarfs increases moving to
later spectral types (or lower masses).  The fastest rotators observed
by these authors in the young disc population (age $\sim 3\ {\rm
  Gyr}$; \citealt{may74}, \citealt{msr91}) were of $\sim$ M4 spectral
type, with $P / \sin i \sim 12\ {\rm hours}$.  Given that we expect
the objects to spin-down on the main sequence, their origin must be
in the most rapid rotators of the present sample ($\sim 0.25\ {\rm
  days}$), suggesting a factor of only $\sim 2$ spin-down.  This is
slower than the \citet{sk72} $t^{1/2}$ prediction of a factor of $\sim
4.5$, suggestive of a very long spin-down timescale for these
late-type dwarfs compared to solar-type dwarfs.

\section{Discussion}
\label{disc_section}

\subsection{Link to angular momentum}

It is instructive to examine the implications of the relations
determined in \S \ref{pmd_section} for the stellar angular momentum
$J$.  We define
\begin{eqnarray}
J &=& I \omega \\
  &=& {2 \pi k^2 M R^2 \over{P}}
\label{j_eqn}
\end{eqnarray}
where $I = k^2 M R^2$ is the moment of inertia of a star of
mass $M$ and radius $R$, $k$ is the radius of gyration (e.g. $k^2 =
2/5$ for a sphere of uniform density), and $\omega = 2 \pi / P$ is the
rotational angular velocity.

\begin{figure}
\centering
\includegraphics[angle=270,width=3in]{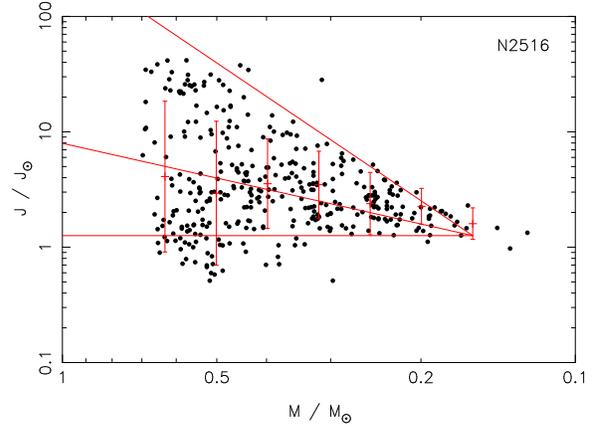}

\caption{Stellar angular momentum $J$ as a function of mass for the
  NGC 2516 rotation period sample.  The lines show $J = {\rm
  constant}$, $J \propto M$ and $J \propto M^3$.  The error bars show
  the median of the distribution binned in $0.1\ {\rm dex}$ bins of
  $\log M$.}

\label{jmd_n2516}
\end{figure}

For a fully-convective star ($M \la 0.4\ \msun$), $k$ is
approximately independent of mass, and therefore the specific angular
momentum $j = J / M \propto R^2 / P$.  Furthermore, on the ZAMS we can
approximate $M \propto R$ at these masses\footnote{Fitting the
  mass-radius relation from the models of \citet{bcah98} over the
  range $0.1 < M/\msun < 0.7$ with a simple linear model gave $R/\rsun
  = 0.8 M/\msun + 0.075$, which is a better approximation than the
  simple proportionality.}  Thus, recalling the simple description of
the morphology of the period versus mass plots in terms of power law
relations from \S \ref{pmd_section} and Figure \ref{pmd}, $P = {\rm
  constant}$ implies $j \propto M^2$, and $P \propto M^3$ implies $j
\propto M^{-1}$.  These correspond to $J \propto M^3$ and $J = {\rm
  constant}$, respectively.    A plot of $J$ as a function of stellar
mass for the NGC 2516 sample is shown in Figure \ref{jmd_n2516}, with
these lines overlaid.

We computed the median rotation period as a function of mass for the
NGC 2516 data, finding it was consistent with $P_{\rm med} \propto
M^2$, a relation inbetween those of the upper and lower envelopes in
Figure \ref{pmd}.  This implies $j = {\rm constant}$ as a
function of mass, a result consistent with the conclusions of
\citet*{h2001} in the ONC.  The lack of any change in the median
relation for stars still undergoing contraction on the PMS indicates
that any mass-dependent angular momentum losses are not important in
determining the evolution of the median over this age range.  However,
this is not the case in detail or once the stars reach the ZAMS, in
particular the {\em shape} of the distribution even in this mass range
does evolve in a mass-dependent fashion.

It is interesting to compare the present observations with the
Pleiades survey of \citet{se2004b}.  The mass-dependence of the
angular momentum evolution we have derived for NGC 2516 appears to be
slightly inconsistent with their conclusions, where they found $P
\propto M$.  Examining the Pleiades diagram in Figure \ref{pmd}
indicates that their result can most likely be attributed to the lack
of data for $0.25 < M/\msun < 0.5$.  Otherwise the results of the
surveys are in good agreement, as expected due to the similar ages of
the two clusters.

The earlier conclusions of \citet{se2004a} in $\sigma$ Ori (age $\sim
3\ {\rm Myr}$) are interesting, since they also found $P \propto M$.
At these ages, the mass-radius relation is better-fit by $R/\rsun \propto
\left(M/\msun\right)^{1/2}$, so this implies $P \propto R^2$,
or constant specific angular momentum, the same as our median relation.

\subsection{Simple models}
\label{model_section}

\subsubsection{Description}

We have developed a simple model scheme, to attempt to find an
explanation for the observations, following the approach of
\citet{bfa97} and \citet{a98}.  In this work, the relation between
$M$, $R$ and $k$ as a function of time for the radiative core (where
present) and convective envelope was taken from the Lyon group
stellar models, generously supplied to us by I. Baraffe (private
communication) for $0.1, 0.2, 0.3, 0.4, 0.6, 0.8$ and $1.0\ \msun$.
We have interpolated the values where necessary for intermediate
masses.

The overall angular momentum loss rate was split into two components:
losses due to stellar winds, assuming a loss law with saturation at a
critical angular velocity $\omega_{\rm sat}$ (allowed to vary as a
function of mass), and losses due to disc locking, which to a good
approximation maintains a constant angular velocity until the
circumstellar disc dissipates.

We assume an angular momentum loss law for stellar winds of the
form suggested by \citet{k88}, modified to include saturation of the
angular momentum losses above a critical angular velocity $\omega_{\rm
  sat}$.  \citet{bs96} argue that saturation at high velocity is
required to account for fast rotators on the ZAMS, and may be
physically motivated by saturation of the magnetic dynamo activity
giving rise to the angular momentum losses.  This is supported by
observations of several diagnostics of magnetic activity, including
saturation of chromospheric emission (e.g. \citealt{vil84}).

The adopted angular momentum loss law has the following form:
\begin{equation}
\left({dJ\over{dt}}\right)_{\rm wind} = \left\{ \begin{array}{r}
-K\ \omega^3\ \left({R \over{R_\odot}}\right)^{1/2} \left({M \over{M_\odot}}\right)^{-1/2}, \omega < \omega_{\rm sat} \\
-K\ \omega\ \omega_{\rm sat}^2\ \left({R \over{R_\odot}}\right)^{1/2} \left({M \over{M_\odot}}\right)^{-1/2}, \omega \ge \omega_{\rm sat}
\end{array} \right.
\label{amloss_eqn}
\end{equation}
where $\omega$ is the angular velocity of the star, $R$ is its radius,
and $M$ its mass.  $K$ is a constant, chosen to give the correct value
of the solar angular momentum for a $1\ \msun$ star at the age of the
Sun.  The value $K = 2.7 \times 10^{47}\ {\rm g\ cm^2\ s}$
was determined by \citet{bfa97} to satisfy this requirement, with
$\omega_{\rm sat} = 14\ \omega_\odot$.  This particular form of
angular momentum loss law is found to reproduce the time dependence of
rotation between the Hyades age ($625\ {\rm Myr}$) and the Sun ($4.57\
{\rm Gyr}$).

The value of $\omega_{\rm sat}$ is found to depend on stellar mass
(e.g. \citealt{k97}, \citealt{bs96}), and is typically assumed to be
inversely proportional to the convective overturn timescale $\tau$:
\begin{equation}
\omega_{\rm sat} = \omega_{\rm sat,\odot}\ {\tau_\odot\over{\tau}}
\label{tauconv_eqn}
\end{equation}
where the quantities subscripted $\odot$ are those for the sun
\citep{k97}.  The values of $\tau$ were taken from the $200\ {\rm
  Myr}$ model of \citet{kd96} for $M > 0.5\ \msun$.  \citet*{spt00}
found that simply linearly extrapolating these models to lower masses
did not work well, and derived a set of empirical values to fit the
distribution seen in the Hyades at these masses.  We have also found
this to be the case for the NGC 2516 data, and adopt the values of
\citet{spt00}, which give a much better fit to the data.

We consider two classes of models: solid body rotation, where the star
is considered to rotate as a single solid body, implying that the core
and envelope must have the same angular velocity, and differential
rotation, where the core and envelope are allowed to decouple, with
angular momentum losses from the envelope, and angular momentum
transfer on a timescale $\tau_c$ between the core and envelope.

In order to account for disc locking, we introduce a disc locking time
$\tau_{\rm disc}$, such that for $t < \tau_{\rm disc}$, the star is
locked to the circumstellar disc, and maintains a constant angular
velocity in the convective envelope, equal to the initial value.  For
$t > \tau_{\rm disc}$, the angular velocity evolves as follows:

For solid body rotation, we can simply differentiate the expression $J
= I \omega$ where $I$, $J$ and $\omega$ are the quantities for the
entire star, with respect to time.  The only angular momentum loss is
due to stellar winds, so:
\begin{equation}
{d\omega\over{dt}} = 
{\omega\over{J}}\ \left({dJ\over{dt}}\right)_{\rm wind} -
{\omega\over{I}}\ {dI\over{dt}}
\end{equation}

For differential rotation, the core and envelope must be considered
separately.  For the envelope, two extra terms are added (see
\citealt{a98} for a full derivation), describing the coupling of
angular momentum from the core (first term), and the loss of mass to
the core as the star evolves (second term): 
\begin{eqnarray*}
{d\omega_{\rm conv}\over{dt}} &=&
{1\over{I_{\rm conv}}}\ {\Delta J\over{\tau_c}}
- {2\over{3}}\ {R_{\rm rad}^2\over{I_{\rm conv}}}\ \omega_{\rm conv}\
{dM_{\rm rad}\over{dt}} \\
&&- {\omega_{\rm conv} \over{J_{\rm conv}}}\
\left({dJ_{\rm conv}\over{dt}}\right)_{\rm wind} -
{\omega_{\rm conv}\over{I_{\rm conv}}}\ {dI_{\rm conv}\over{dt}}
\label{omegaconv_eq}
\end{eqnarray*}
where we assume:
\begin{equation}
\Delta J = {I_{\rm conv} J_{\rm rad} - I_{\rm rad} J_{\rm
    conv}\over{I_{\rm rad} + I_{\rm conv}}}
\end{equation}
as suggested by \citet{mac91}.

The corresponding equation for the evolution of the core angular
velocity is:
\begin{equation}
{d\omega_{\rm rad}\over{dt}} =
-{1\over{I_{\rm rad}}}\ {\Delta J\over{\tau_c}}
+ {2\over{3}}\ {R_{\rm rad}^2\over{I_{\rm rad}}}\ \omega_{\rm conv}\
{dM_{\rm rad}\over{dt}}
- {\omega_{\rm rad}\over{I_{\rm rad}}}\ {dI_{\rm rad}\over{dt}}
\label{omegarad_eq}
\end{equation}
where the first two terms again represent coupling of angular momentum
to the envelope, and mass gained by the core (from the convective
envelope), respectively.  The final term accounts for the change in
moment of inertia of the core as the star evolves.

\subsubsection{Evolution from NGC 2362}
\label{ove_section}

We first attempt to evolve the observed rotation rates from our survey
in NGC 2362 (Hodgkin et al., in prep) forward in time to reproduce the
available rotation period data in NGC 2516.  In order to do this, we
consider the slowest rotators, characterised by the $25$th percentile
(the lower quartile) of the distribution of observed angular
velocities, $\omega$, and the fastest rotators, characterised by the
$90$th percentile of the distribution.  The discrepancy in measures
here is because the $10$ percentile was found to be somewhat unstable
in the presence of small number statistics for slow rotators in some
of the clusters under consideration.  Nevertheless, the results for
this quantity are compatible with the conclusions we draw for the $25$
percentile for those clusters with sufficiently good statistics.

We fit solid body and differentially rotating models to the data,
allowing the parameters $\omega_{\rm sat}$, $\tau_{\rm disc}$ and
$\tau_c$ to vary to fit the slowest rotators ($25$ percentile).

\begin{figure*}
\centering
\includegraphics[angle=0,width=6in]{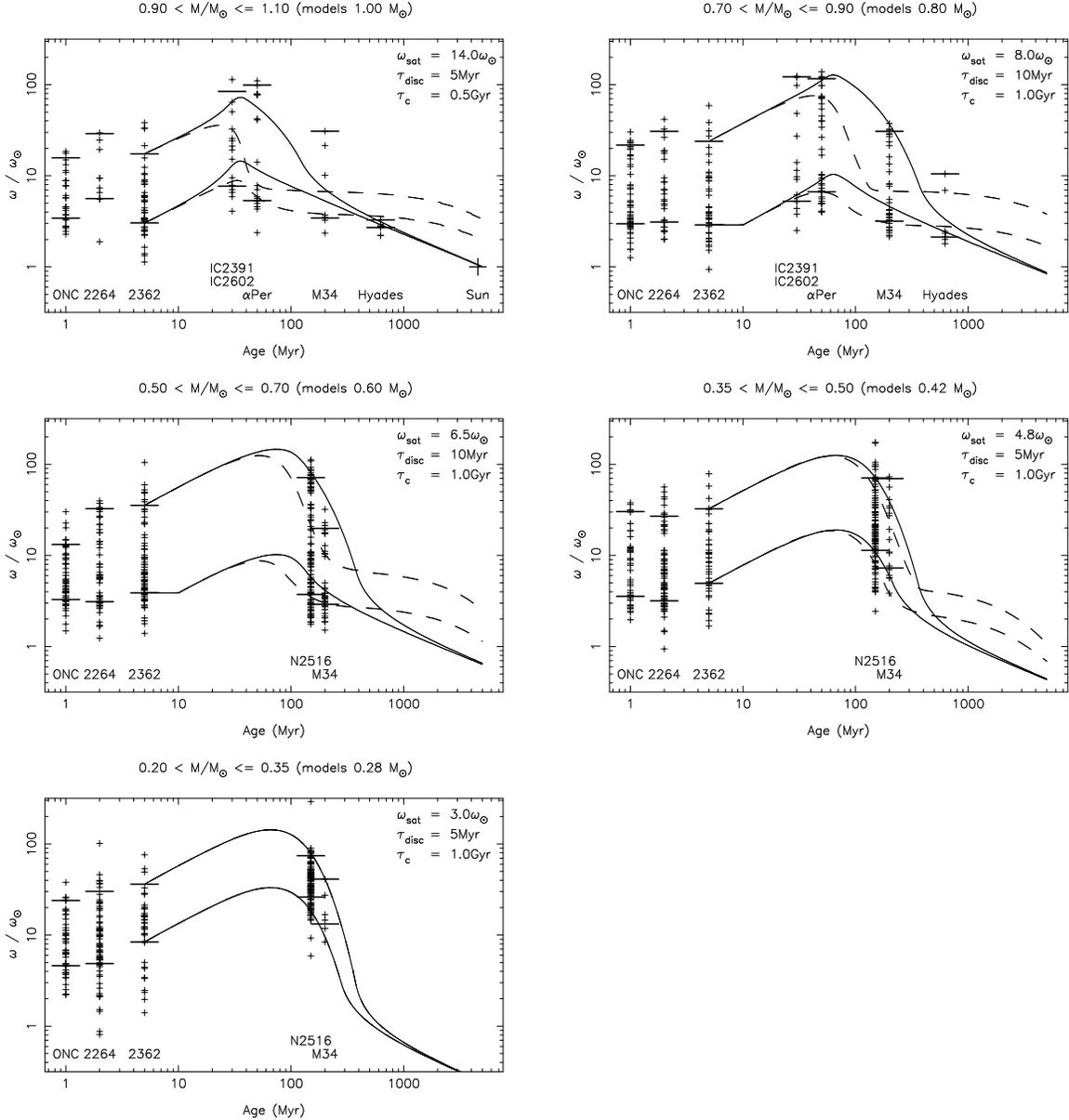}

\caption{Rotational angular velocity $\omega$ plotted as a function of
  time in five mass bins: $0.9 < M/\msun \le 1.1$, $0.7 < M/\msun \le
  0.9$, $0.5 < M/\msun \le 0.7$, $0.35 < M/\msun \le 0.5$, and $0.2 <
  M/\msun \le 0.35$.  Crosses show the rotation period data, and short
  horizontal lines the $25$th and $90$th percentiles of $\omega$, used
  to characterise the slow and fast rotators respectively.
  The lines show our models for $1.0$, $0.8$, $0.6$, $0.42$ and $0.28\
  \msun$ (respectively), where the solid lines are solid body models,
  and dashed lines are differentially rotating models, with the
  parameters shown.  Plotted are the ONC ($1\ {\rm Myr}$), NGC 2264
  ($2\ {\rm Myr}$), NGC 2362 ($5\ {\rm Myr}$), IC 2391, IC 2602 ($\sim
  30\ {\rm Myr}$), $\alpha$ Per ($\sim 50\ {\rm Myr}$), NGC 2516, M34,
  the Hyades ($625\ {\rm Myr}$) and the Sun ($\sim 4.57\ {\rm Gyr}$).
  The IC 2391 data were taken from \citet{ps96} and IC 2602 from
  \citet{bsps99}.  The $\alpha$ Per data are a compilation of the
  results  from \citet{aper1}, \citet{aper2}, \citet{ple4},
  \citet{aper3}, \citet{ple5}, \citet{aper4}, \citet{ple6},
  \citet{aper5}, \citet{aper6}, \citet{aper7}, \citet{aper8},
  \citet{aper9}, \citet{aper10}, \citet{aper11}, and the Hyades data
  from \citet{hya1} and \citet{ple6}, taken from the open cluster
  database.}
\label{pevol}
\end{figure*}

Figure \ref{pevol} shows our results, where we performed the analysis
in a series of mass bins, in order to resolve the mass dependence of
the evolution.  These were: $0.9 < M/\msun \le 1.1$, $0.7 < M/\msun
\le 0.9$, $0.5 < M/\msun \le 0.7$, $0.35 < M/\msun \le 0.5$, and $0.2
< M/\msun \le 0.35$ (chosen empirically to encompass the changes in
behaviour seen in the models, while retaining reasonable statistics).
The models were calculated for single masses roughly at the centre of
these bins, of $1.0$, $0.8$, $0.6$, $0.42$ and $0.28\ \msun$.

The solid body models shown in Figure \ref{pevol} show good overall
agreement with the upper $90$ percentiles for the NGC 2516 data, in
all the mass bins.  For the $0.9 < M/\msun < 1.1$ bin, the M34 upper
$90$ percentile is at much greater $\omega$ than the model, but we
suspect that this result is not significant due to small number
statistics in the M34 sample at these masses: the discrepancy is due
to the presence of $3$ data points, and as discussed in \citet{i2006},
these may be contaminants.  For $0.7 < M/\msun < 0.9$ the upper $90$
percentile in the Hyades appears to lie at much greater $\omega$ than
the model, but again this result is strongly affected by small number
statistics.  In order to resolve these issues, improved statistics at
these masses and ages will be required.  We note that the
differentially rotating models with the assumed values of $\tau_c$ give
a poor fit to the fast rotators (as characterised by the $90$
percentiles), especially on the ZAMS.

For the lower $25$ percentiles, the solid body models give a poor fit
to the evolution from NGC 2362 to older ages, which becomes
particularly apparent after $\sim 50\ {\rm Myr}$ to the age of the
Hyades in the top two mass bins.  Introducing core-envelope decoupling
with $\tau_c = 1.0\ {\rm Gyr}$ provides a much better fit to the
observations, and produces curves in good agreement with all the
presently available data.  We note however that we have not
constrained these to fit the sun, which requires a modification of the
constant $K$ in Eq. (\ref{amloss_eqn}).

An interesting implication of the results in Figure \ref{pevol} is
that we cannot find a model in this scheme that fits both the slow and
fast rotators simultaneously: the values of $\tau_c$ required in the
models are different (solid body models correspond to $\tau_c = 0$).
The result that the fast rotators seem to behave more like solid
bodies may indicate that the redistribution of angular momentum at the
core/envelope boundary is more efficient in fast rotators.

We note that there are relatively few constraints on the evolution for
masses $< 0.7\ \msun$ at the intermediate ages ($5$ to $150\ {\rm
  Myr}$), and at old ages ($> 200\ {\rm Myr}$).  We have recently
obtained data in NGC 2547 ($\sim 38.5\ {\rm Myr}$; \citealt{nj2006})
at these masses, which should allow us to address the first of these
problems in a forthcoming publication.

\subsubsection{Constraints on initial conditions}

\begin{figure}
\centering
\includegraphics[angle=270,width=3in]{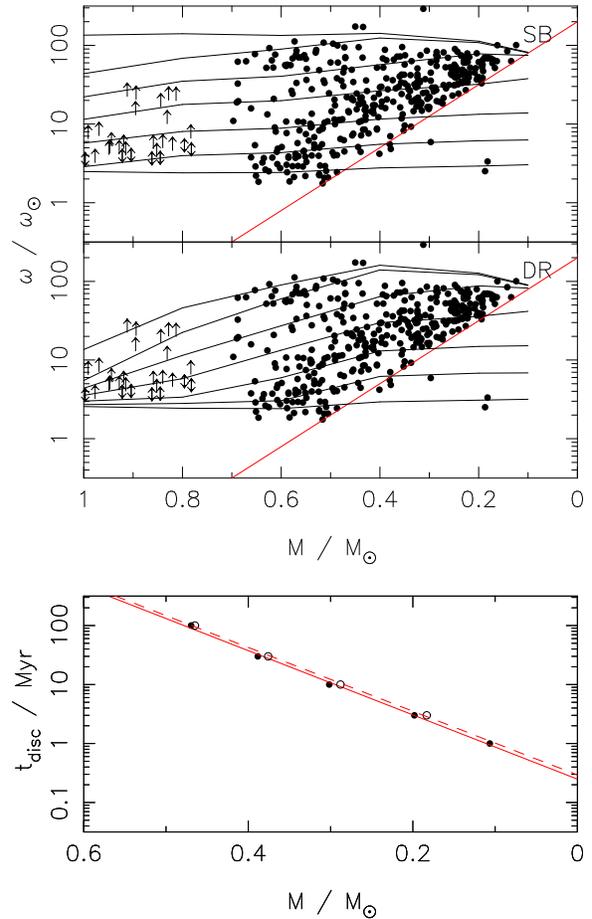}

\caption{{\bf Top and centre:} NGC 2516 data (points), plus lower
  limits derived from the $v \sin i$ values of \citet{ter2002}
  (arrows), with models overlaid (top: solid body model, centre:
  differentially rotating model) for $t_{\rm disc} = 0$, $0.3$, $1$,
  $3$, $10$, $30$ and $100\ {\rm Myr}$ (top to bottom).  The initial
  period $P_0$ was $10\ {\rm days}$.  The straight line shows the
  approximate lower envelope used to derive the disc lifetime as a
  function of mass.  {\bf Bottom:} derived disc lifetime $t_{\rm
  disc}$ as a function of mass, for the solid body models (filled
  circles and solid line), and differentially rotating models (open
  circles and dashed line).  The models used the $\omega_{\rm sat}$
  values of \citet{spt00} (these were adjusted to provide a good fit
  to the Hyades data).}
\label{pevol_tdisc}
\end{figure}

We explore here an alternative method for analysis of the NGC 2516
data, following \citet{spt00}.  By assuming that all stars share the
same initial rotation rate, here $10$ days, as used by \citet{spt00},
we can constrain the disc lifetimes required to reproduce the observed
envelope in diagrams of $\omega$ as a function of stellar mass.  In
particular, the observed lower envelope as a function of mass (the
slowest rotators seen) must result from the longest disc lifetimes,
and hence we can derive the required maximum disc lifetime as a
function of mass.  Figure \ref{pevol_tdisc} shows the results of this
analysis for NGC 2516.

The results in Figure \ref{pevol_tdisc} indicate that the disc
lifetime must decrease strongly as a function of decreasing mass in
order to reproduce the NGC 2516 data with a single initial rotation
period.  Lifetimes of $> 10\ {\rm Myr}$ are required for $M > 0.3\
\msun$, a result at odds with other measurements
(e.g. \citealt{hll2001}; \citealt{b2006} and references therein), which
indicate values of $\sim 10\ {\rm Myr}$ at solar mass.  Therefore it
seems unlikely that the envelope shape we observe is due purely to
varying disc lifetimes as a function of mass.

\begin{figure}
\centering
\includegraphics[angle=270,width=3in]{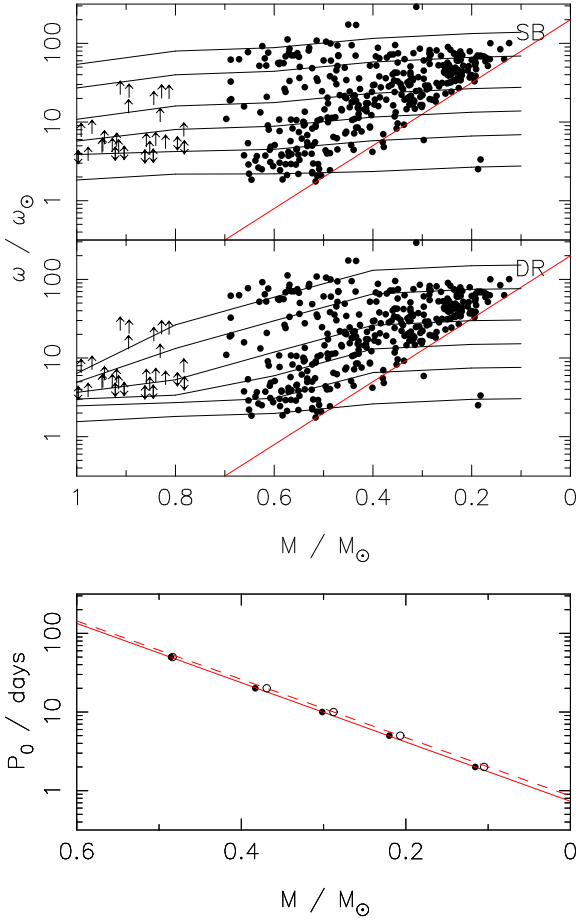}

\caption{As Figure \ref{pevol_tdisc}, but for varying $P_0$, with
  fixed $t_{\rm disc} = 10\ {\rm Myr}$.  The curves show the models
  for $P_0 = 1$, $2$, $5$, $10$, $20$, $50\ {\rm days}$ (top to
  bottom).}
\label{pevol_p0}
\end{figure}

An alternative explanation (the other extreme) is that the initial
period $P_0$ varies as a function of mass, but the disc lifetime
$t_{\rm disc}$ remains fixed.  Figure \ref{pevol_p0} shows the results
of doing this for the NGC 2516 data, assuming $t_{\rm disc} = 10\ {\rm
  Myr}$, indicating again that a rather wide range of initial
conditions appear to be required to explain the observed envelope
purely by variations of $P_0$.  Furthermore, the very slow rotation
($P_0 \sim 60\ {\rm days}$) apparently required at $0.5\ \msun$ is
somewhat implausible in light of measurements of very young
populations, where typical rotation periods for stars of this mass are
$\sim 10\ {\rm days}$.

In reality, a combination of the two effects is likely, which will
lead to intermediate values of $t_{\rm disc}$ and $P_0$.  However, the
required parameter ranges remain implausible in light of existing
data.  It is possible that improved modelling of the star using a more
complete treatment of differential rotation (e.g. \citealt{spt00}) will
solve this issue.

\subsubsection{Value of $\omega_{\rm sat}$}

\begin{figure}
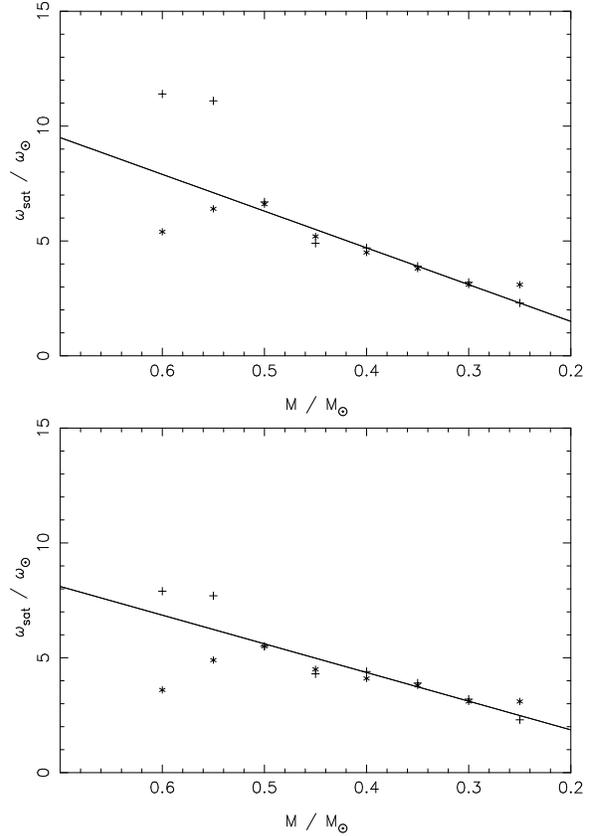

\centering
\includegraphics[angle=270,width=3in]{omegasat.ps}
\includegraphics[angle=270,width=3in]{omegasat_decoup.ps}

\caption{$\omega_{\rm sat}$ plotted as a function of mass, in $0.1\
  \msun$ bins, from fitting to the NGC 2516 $j$ distribution lower 25
  percentile (plus symbols) and upper 90 percentile (stars), using NGC
  2362 as the initial condition, for solid body models (top) and
  differentially rotating models (bottom; $\tau_c = 1.0\ {\rm Gyr}$).
  This diagram indicates that a single functional form fails to fit
  all of these measures simultaneously for $M \ga 0.5\ \msun$, but
  that a simple linear relation is a good approximation at lower
  masses.  The straight lines show a fit for $M \le 0.5\ \msun$,
  finding $(\omega_{\rm sat} / \omega_\odot) = 16.0\ (M / \msun) -
  1.7$ (solid body models) and $(\omega_{\rm sat} / \omega_\odot) =
  12.5\ (M / \msun) - 0.6$ (differentially rotating models).}
  
\label{omegasat_figure}
\end{figure}

The value of $\omega_{\rm sat}$ at each mass is tightly constrained
by the need to reproduce the NGC 2516, M34 and Hyades data.  In
\ref{ove_section}, we derived values for $\omega_{\rm sat}$ in coarse
mass bins, to produce model curves providing a reasonable explanation
for the observations.  In order to investigate the mass dependence of
this parameter in detail, we used the NGC 2362 and NGC 2516
observations to perform a fit of $\omega_{\rm sat}$ as a function
of mass in much smaller bins to give better sampling, using the same
percentile based technique.  Figure \ref{omegasat_figure} shows the
results, finding a linear dependence of $\omega_{\rm sat}$ on mass to
be a good approximation at the lowest masses, for both solid body and
differentially rotating models.  There is some evidence
that the relation does not apply for higher masses ($M \ga 0.5\
\msun$), but in reality $\omega_{\rm sat}$ is not well-constrained
here, since the majority of these stars (which are slow rotators)
only reach the saturation velocity for a short time during the
evolution from $5$ to $150\ {\rm Myr}$.

From the fit, at solar mass,
$0.8\ \msun$ and $0.5\ \msun$, the relation gives $\omega_{\rm sat} =
14.3, 11.1, 6.3\ \omega_\odot$ respectively for the solid body models,
and $\omega_{\rm sat} = 11.9, 9.35, 5.61\ \omega_\odot$ for the
differentially rotating models.  The value for solar mass and solid
body rotation is very close to the value of $14\ \omega_\odot$ we
assumed earlier, as derived from the convective turnover timescale,
but at $0.8$ and $0.5\ \msun$, $\omega_{\rm sat}$ is larger than the
values of $8\ \omega_\odot$ and $3\ \omega_\odot$ predicted by
Eq. (\ref{tauconv_eqn}): the slope of $\omega_{\rm sat}$ as a function
of mass is less steep than predicted from the convective turnover
timescale.

The decrease of $\omega_{\rm sat}$ with decreasing mass, as determined
in this work, effectively reduces the amount of angular momentum loss
experienced by lower mass stars, compared to what would be experienced
if the value was the same as at solar mass.  This has also been seen
in previous work, e.g. \citet{bs96}.

\subsubsection{Detailed evolution}

\begin{figure}
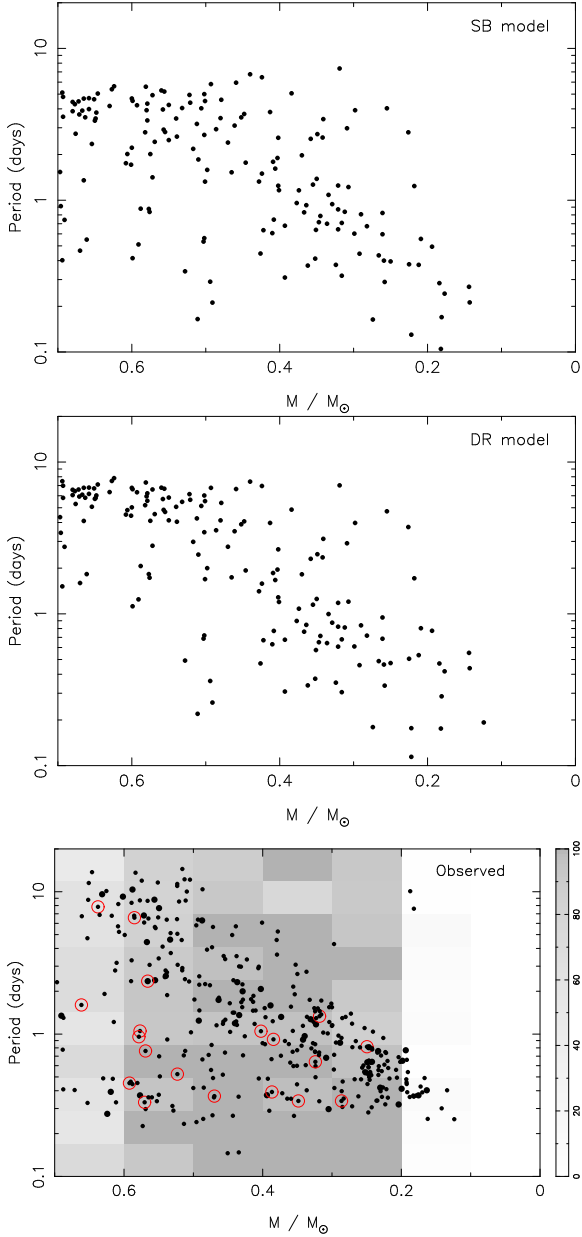

\centering
\includegraphics[angle=270,width=3in]{pmd_n2362_evol_n2516.ps}
\includegraphics[angle=270,width=3in]{pmd_n2362_evol_n2516_decoup.ps}
\includegraphics[angle=270,width=3in]{pmd_n2516_evolcomp.ps}

\caption{Rotation period as a function of mass, using the model
  presented in \S \ref{model_section} to evolve the NGC 2362
  distribution of Figure \ref{pmd} forward in time from $5\ {\rm
  Myr}$ to $150\ {\rm Myr}$ for the solid body (upper panel) and
  differentially rotating (middle panel) models, and the observed NGC
  2516 distribution for comparison (lower panel).}

\label{jmd_n2362_evol}
\end{figure}

Figure \ref{jmd_n2362_evol} shows the results of attempting to evolve
the measured rotation periods in NGC 2362 forward in time to the age
of NGC 2516 using the models we have described, with the relations fit
from Figure \ref{omegasat_figure} for $\omega_{\rm sat}$ as a function
of mass, on an object-by-object basis.  By doing this, we can test if
the model we have presented can reproduce the observations of NGC
2516, given the NGC 2362 rotators as in input.  Note that we have
neglected disc locking in this analysis for simplicity.

The apparent mass dependent morphology of the NGC 2516 distribution is 
well-reproduced from the NGC 2362 distribution by the models, with two
exceptions: for the slow rotators at the highest masses in the
diagram, the predicted periods are too short by a factor of $\sim 2-3$
compared to the observations for the solid body models, and slightly
too short for the differentially rotating models.  This suggests that
slow rotators experience more angular momentum loss on the ZAMS
than the models predict.  Conversely, the solid body models provide a
better fit to the fast rotators, especially in the mass range $0.6 -
0.8\ \msun$.  This is not surprising in light of the conclusions of \S
\ref{ove_section}.

In particular, the form of mass dependence for $M \la 0.6\ \msun$ is
approximately correct, with the exception of a small number of objects
rotating faster than the apparent `limit' of $\sim 0.2\ {\rm days}$
seen in the NGC 2516 data.  A small number of objects are also visible
in the evolved NGC 2362 distribution above the $P \propto M^3$
relation seen for the slowest rotators in NGC 2516.  Since these
originated as relatively slow rotators in NGC 2362, we suspect that
they are the result of field contamination, or errors in the measured
rotation periods in that sample due to the limited time baseline of
the observations.

This work suggests that the mass dependence of the observed rotation
period distribution is reproduced by the combination of the initial
conditions as seen in NGC 2362, the mass-dependence of stellar
contraction on the PMS, and the mass-dependence of the saturation
angular velocity $\omega_{\rm sat}$.  Furthermore, the convergence in
rotation rates as a function of time (see also Figure \ref{pevol}) is
explained reasonably well by saturation of the angular momentum
losses, over the mass range of interest for NGC 2516, apparently
without need to invoke any further effects.

\begin{figure}
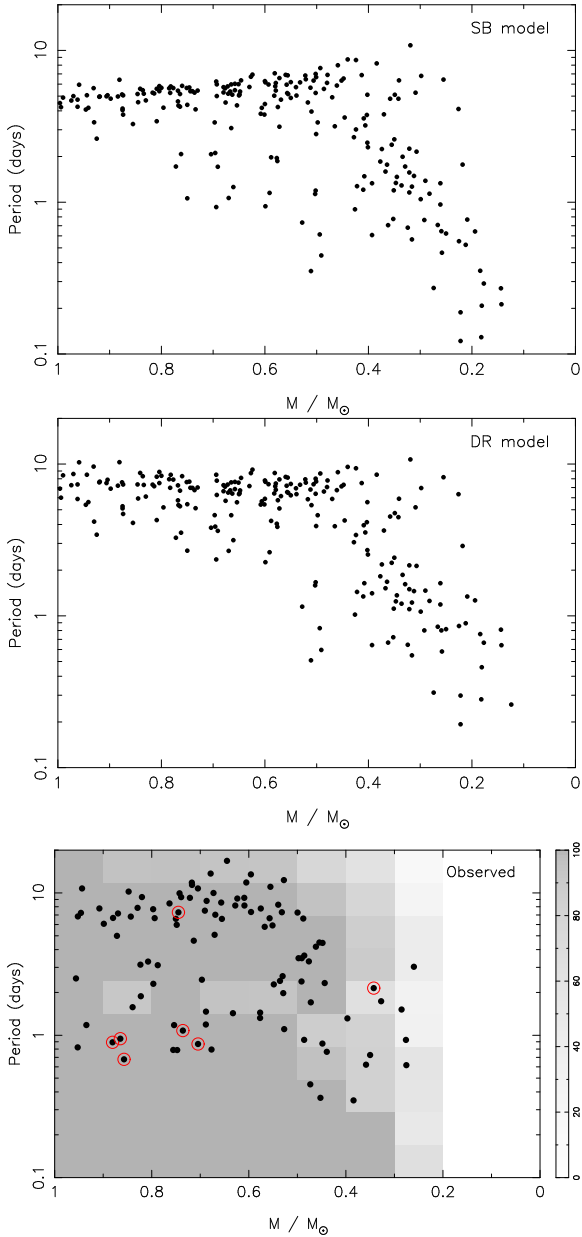

\centering
\includegraphics[angle=270,width=3in]{pmd_n2362_evol_m34.ps}
\includegraphics[angle=270,width=3in]{pmd_n2362_evol_m34_decoup.ps}
\includegraphics[angle=270,width=3in]{pmd_m34.ps}

\caption{Rotation period as a function of mass, using the model
  presented in \S \ref{model_section} to evolve the NGC 2362
  distribution of Figure \ref{pmd} forward in time from $5\ {\rm
  Myr}$ to $200\ {\rm Myr}$ for the solid body (upper panel) and
  differentially rotating (middle panel) models, and the observed M34 
  distribution for comparison (lower panel).}

\label{jmd_n2362_evol_m34}
\end{figure}

Figure \ref{jmd_n2362_evol_m34} shows the same comparison for M34,
showing that the feature of constant rotation period as a function of
mass for $M \ga 0.6\ \msun$ is also well-reproduced by the models, at
approximately the correct period in the case of the differentially
rotating models, and again at a shorter period than the observed value
for the solid body models.  It should be noted that the adopted values
of $\omega_{\rm sat}$ in this mass range are not well-constrained by
the NGC 2516 data used to calibrate them, and have a strong effect on
the morphology of this part of the diagram, so it is possible that any
deviations here result from our assumption of a linear relation over
the entire mass range.

\section{Conclusions}
\label{conclusions_section}

We have reported on results of an $i$-band photometric survey of NGC
2516, covering $\sim 1\ {\rm sq. deg}$ of the cluster.  Selection
of candidate members in a $V$ versus $V-I$ colour-magnitude diagram
using an empirical fit to the cluster main sequence found $1685$
candidate members, over a $V$ magnitude range of $16 < V < 26$
(covering masses from $0.7\ \msun$ down to below the brown dwarf
limit).  The likely field contamination level was estimated using a
simulated catalogue of field objects from the Besan\c{c}on Galactic
models \citep{r2003}, finding that $\sim 650$ objects were likely
field contaminants, an overall contamination level of $\sim 39 \%$,
implying that there are $\sim 1000$ real cluster members over this
mass range in our field-of-view.

From $\sim 8\ {\rm nights}$ of time-series photometry we derived
lightcurves for $\sim 104\,000$ objects in the NGC 2516 field,
achieving a precision of $< 1 \%$ per data point over $15 \la i \la 19$. 
The lightcurves of our candidate cluster members were searched for
periodic variability corresponding to stellar rotation, giving $362$
detections over the mass range $0.15 < M/\msun < 0.7$.

A striking morphology was found in the rotation period distribution as
a function of both mass, with a median relation of $P \propto M^2$ (or
constant specific angular momentum $j$) for masses below $\sim 0.6\
\msun$, with the slowest rotators (the majority of the sample)
following $P \propto M^3$ ($j \propto M^{-1}$), with the fastest
rotators over the entire mass range having $P \ga 0.25\ {\rm days}$
(corresponding to a line of $j \propto M^2$).  Our earlier M34 results
indicate that for $M \ga 0.6\ \msun$, the slowest rotators follow
instead $P \sim {\rm constant}$ as a function of mass, with a
transition between the two regimes at $M \sim 0.6\ \msun$.  We suggest
that this mass corresponds to the transition to convectively-dominated
stellar interiors moving to lower masses.  This could be tested by
examining rotation period distributions over a wider range of ages for
M-dwarfs.

In \S \ref{model_section}, simple models of the rotational evolution
were considered, both for the solid body case, and including
differential rotation between a decoupled radiative core and convective
envelope, following \citet{bfa97} and \citet{a98}.  A \citet{k88}
angular momentum loss law incorporating saturation at a critical
angular velocity $\omega_{\rm sat}$ was assumed.  The results
indicate that a mass-dependent saturation velocity (e.g. of the form
suggested by \citealt{spt00}) is required to reproduce the
observations.  Furthermore, we find that solid body rotation is
sufficient to explain the evolution of the fastest rotators, but
core-envelope decoupling appears to be needed to explain the evolution
of the slowest rotators.  This may indicate that the redistribution of
angular momentum at the core/envelope boundary is  more efficient in
fast rotators.  We were unable to find a model which fits both the
fast and slow rotators simultaneously.  Further rotation period
measurements at low masses, particularly for ages intermediate between
$5$ and $150\ {\rm Myr}$ and for ages $> 200\ {\rm Myr}$, will provide
more stringent constraints on the models, but the present observations
already indicate the need for more theoretical work.

We intend to publish the final catalogue of NGC 2516 membership
candidates after obtaining follow-up spectroscopy.  However, the
preliminary catalogue of membership candidates is available on
request.

\section*{Acknowledgments}

Based on observations obtained at Cerro Tololo Inter-American
Observatory, a division of the National Optical Astronomy
Observatories, which is operated by the Association of Universities
for Research in Astronomy, Inc. under cooperative agreement with the
National Science Foundation.  This publication makes use of data
products from the Two Micron All Sky Survey, which is a joint project
of the University of Massachusetts and the Infrared Processing and
Analysis Center/California Institute of Technology, funded by the
National Aeronautics and Space Administration and the National Science
Foundation.  This research has also made use of the SIMBAD database,
operated at CDS, Strasbourg, France.  The Open Cluster Database, as
provided by C.F. Prosser and J.R. Stauffer, may currently
be accessed at {\tt http://www.noao.edu/noao/staff/cprosser/}, or by
anonymous ftp to {\tt 140.252.1.11}, {\tt cd /pub/prosser/clusters/}.

JI gratefully acknowledges the support of a PPARC studentship, and SA
the support of a PPARC postdoctoral fellowship.  We would like to
express our gratitude to Isabelle Baraffe for providing the stellar
evolution model tracks used in \S \ref{model_section}, and to the
anonymous referee for comments that have helped to improve the paper.

\end{document}